\documentclass[prd,aps,showpacs,superscriptaddress,floatfix,preprintnumbers,reprint,nofootinbib]
{
revtex4-1}
\usepackage{color}
\usepackage{epsfig}
\usepackage{amsmath,bm}
\usepackage{blindtext}
\usepackage[modulo]{lineno}
\usepackage{xspace}
\usepackage{enumitem}
\usepackage[normalem]{ulem} %provides \sout{} to strike-out text

\newcommand{\rmL}{\ensuremath{\mathrm L}}
\newcommand{\rmZ}{\ensuremath{\mathrm Z}}
\newcommand{\rmE}{\ensuremath{\mathrm e}}
\newcommand{\rmV}{\ensuremath{\mathrm V}}
\newcommand{\rmA}{\ensuremath{\mathrm A}}
\newcommand{\der}{\ensuremath{\,\mathrm{d}}}
\newcommand{\GeV}{\ensuremath{\mathrm{GeV}}}

\newcommand{\bsym}[1]{\ensuremath{\boldsymbol{#1}}}
\newcommand{\ve}[1]{\bsym{#1}}

\newcommand{\msbar}{\ensuremath{\overline{\mathrm MS}}}
\newcommand{\gammaZ}{\ensuremath{\gamma\rmZ}}

\newcommand{\Sup}[2]{\ensuremath{#1^{\mathrm{#2}}}}

\newcommand{\ppt}[1]{\ensuremath{\tilde{p}_{\mathrm{#1}}}}
\newcommand{\fpt}[1]{\ensuremath{f_{\mathrm{#1}}}}
\newcommand{\tht}[1]{\ensuremath{\theta_{\mathrm{#1}}}}
\newcommand{\epmp}{\ensuremath{\mathrm{e}^{\pm}\mathrm{p}}\xspace}
\newcommand{\eplp}{\ensuremath{\mathrm{e}^{+}\mathrm{p}}}
\newcommand{\emip}{\ensuremath{\mathrm{e}^{-}\mathrm{p}}}
\newcommand{\Fg}[1]{\ensuremath{F_{#1}^{\gamma}}}
\newcommand{\FgZ}[1]{\ensuremath{F_{#1}^{\gammaZ}}}
\newcommand{\FZ}[1]{\ensuremath{F_{#1}^{\rmZ}}}
\newcommand{\eFL}{\ensuremath{F_{\rmL}}}
\newcommand{\kaZ}{\ensuremath{k_{\rmZ}}}
\newcommand{\vve}{\ensuremath{v_{\rmE}}}
\newcommand{\aae}{\ensuremath{a_{\rmE}}}
\newcommand{\emz}{\ensuremath{m_{\rmZ}}}
\newcommand{\ggV}{\ensuremath{g_{\rmV}}}
\newcommand{\ggA}{\ensuremath{g_{\rmA}}}

\newcommand{\lamvg}{\ensuremath{\Sup{\lambda}{v}_g}}
\newcommand{\lamsg}{\ensuremath{\Sup{\lambda}{s}_g}}
\newcommand{\lamqb}{\ensuremath{\lambda_{\bar{q}}}} 
\newcommand{\Kqb}{\ensuremath{K_{\bar{q}}}}

\newcommand{\xsc}{\textsc}
\newcommand{\xtt}{\texttt}
\newcommand{\xit}{\textit}
\newcommand{\xbf}{\textbf}

\newcommand{\qcdnum}{\xsc{qcdnum}\xspace}
\newcommand{\splint}{\xsc{splint}\xspace}
\newcommand{\bat}{\xsc{bat}\xspace}
\newcommand{\julia}{\xsc{julia}\xspace}
\newcommand{\fortranzz}{\xsc{fortran}{\footnotesize 77}\xspace}

\newcommand{\batjl}{\xtt{BAT.jl}\xspace}
\newcommand{\partondensity}{\xtt{PartonDensity.jl}\xspace}
\newcommand{\qcdnumjl}{\xtt{QCDNUM.jl}\xspace}
\newcommand{\binarybuilderjl}{\xtt{BinaryBuilder.jl}\xspace}

\newcommand{\pdf}[1][]{PDF#1\xspace}       %use in the middle of line
\newcommand{\secref}[1]{Section~\ref{#1}}
\newcommand{\secs}[2]{Sections~\ref{#1} and~\ref{#2}}
\newcommand{\tabref}[1]{Table~\ref{#1}}

\newcommand{\fig}[1]{Fig.~\ref{#1}}

\newcommand{\Eq}[1]{Eq.~\ref{#1}}
\newcommand{\Eqs}[2]{Eqs.~\ref{#1} and~\ref{#2}}
\newcommand{\Fig}[1]{Fig.~\ref{#1}}

\newcommand{\mbstrut}[1]{\rule{0mm}{#1}}

      \makeatletter
      \newcommand*{\textalltt}{}
      \DeclareRobustCommand*{\textalltt}{%
	      \begingroup
	      \let\do\@makeother
	      \dospecials
	      \catcode`\\=\z@
	      \catcode`\{=\@ne
	      \catcode`\}=\tw@
	      \verbatim@font\@noligs
	      \@vobeyspaces
	      \frenchspacing
	      \@textalltt
      }
      \newcommand*{\@textalltt}[1]{%
	      #1%
	      \endgroup
      }
      \makeatother

\nolinenumbers

\begin{document}

\title{PartonDensity.jl: a novel parton density determination code}

\author{Francesca Capel}\email{capel@mpp.mpg.de}
\affiliation{Max-Planck-Institut f\"ur Physik, M\"unchen, Germany}

\author{Ritu Aggarwal}\email{ritu.aggarwal1@gmail.com}
\affiliation{USAR, Guru Gobind Singh Indraprastha University, East Delhi-110032}

\author{Michiel Botje}\email{m.botje@nikhef.nl}
\affiliation{Nikhef, Amsterdam, The Netherlands}

\author{Allen Caldwell}\email{caldwell@mpp.mpg.de}
\affiliation{Max-Planck-Institut f\"ur Physik, M\"unchen, Germany}

\author{Oliver Schulz}\email{oschulz@mpp.mpg.de}
\affiliation{Max-Planck-Institut f\"ur Physik, M\"unchen, Germany}

\author{Andrii Verbytskyi}\email{andrii.verbytskyi@mpp.mpg.de}
\affiliation{Max-Planck-Institut f\"ur Physik, M\"unchen, Germany}

\begin{abstract}
We introduce our novel Bayesian parton density determination code, \partondensity.  The motivation for this new code, the framework and its validation are described. As we show, \partondensity provides both a flexible environment for the determination of parton densities and a wealth of information concerning the knowledge update provided by the analyzed data set.
\end{abstract}

\maketitle

\section{Introduction}

This paper describes a novel approach to the determination of parton density functions
(\pdf[s]) of hadrons and presents the \partondensity programming code that implements
this approach.  

Our analysis method significantly differs from those pursued by other 
groups~\cite{Cridge:2021pxm,Bailey:2020ooq,Hou:2019efy,Dulat:2015mca,Alekhin:2018pai,
Alekhin:2013nda,NNPDF:2021njg,NNPDF:2017mvq,H1:2021xxi,H1:2015ubc}, allowing us to tackle
data sets that have not been included in \pdf determinations so far. There are two distinct features that set our code apart, as described in the following.

First of all, we use Bayesian techniques to determine the parton density functions.\footnote{
  For other efforts  towards \pdf determination from a Bayesian perspective, see 
  \xit{e.g.}~\cite{Cocuzza:2021cbi,Hunt-Smith:2022ugn}.
}
This formulation allows us to include in a coherent and transparent way known constraints
and results from previous theoretical and experimental analyses in prior probability
distributions. The result of a Bayesian analysis is a multivariate posterior probability
distribution of all model parameters, including nuisance parameters used to describe
systematic uncertainties in the data.  Correlations between any subset of parameters can
be studied, providing vastly more information than results obtained otherwise.
Systematic error propagation is simply achieved by integrating the posterior over the
nuisance parameters. In addition, the information content of the data is easily judged
by comparing the posterior to the input priors. 
    
Secondly, we have implemented a forward modeling approach, where
event numbers in kinematic bins are predicted and compared to observed event counts, which are always bin-by-bin uncorrelated with known Poisson distribution. 
This is in contrast to the analyses of unfolded cross
sections, where a Gaussian statistical hypothesis is implied and where the data are
always correlated in a way that is often not known. 
Another advantage of forward modeling is that it can correctly handle
the low statistics case with empty or sparsely populated bins.
    
In principle, the Bayesian and forward modeling approaches can be used independently of each other. The current version of our code allows for the analysis of inclusive \epmp scattering data,
made available in the form of binned event counts.
An extension to allow for the Bayesian analysis of differential cross sections extracted at the QED-Born
level is in development and will be reported separately. We note that it is also possible to use the forward modeling approach within any likelihood-based analysis framework, not just the Bayesian formalism described here. 

The \partondensity software package\footnote{
  Available at \href{https://github.com/cescalara/PartonDensity.jl}
  {https://github.com/cescalara/PartonDensity.jl}.
}
is implemented in the modern \julia~\cite{ref:Julia} language and uses several \julia
packages for the analysis, the most important of which is the Bayesian Analysis
Toolkit (\batjl)~\cite{Schulz:2021BAT}. We have published an analysis of the ZEUS data~\cite{ZEUS:2013szj,ZEUS:2020ddd} using \partondensity in~\cite{Aggarwal:2022cki}. In this work, we present our analysis method in more detail and demonstrate its validity through the application to relevant simulated pseudo-data, for which the true \pdf[s] are known.  

This paper is organized as follows. In \secref{bayes}, we
begin with a general introduction to the Bayesian analysis approach.
In \secref{se:epAnalysis}, we describe our forward model, or how the \epmp\ binned event counts
are computed from a set of parameterized input \pdf[s].
In the development of the code, a large amount of effort was dedicated to speeding up
the necessary calculations, and we briefly report on these in ~\secref{se:technical}. In \secref{sec:PDFs}, we discuss our \pdf parameterizations
and prior parameter constraints. Finally, validation tests on pseudo-data
and goodness-of-fit tests are described in \secs{se:validation}{se:GOFtests},
respectively. 

\section{Bayesian Analysis Approach \label{bayes}}

\subsection{Introduction}

We use a Bayesian approach to obtain the joint posterior distribution 
$p(\theta|D,M)$ of a set of model parameters~$\theta$, conditional to 
the data set $D$ being analyzed. It is also conditional to the particular
model choices
$M$ such as, among others, the modeling of systematic
uncertainties in the data and the specific form of the \pdf[s] chosen. 

Because it is not possible to obtain an analytic expression for the posterior,
Monte Carlo techniques are used to create parameter samples that are distributed
according to~$p(\theta|D,M)$. Nuisance parameters that are needed in the modeling
of the data but are of no interest are removed by integrating the posterior
over these. In this way, 
systematic uncertainties are automatically propagated in a
consistent fashion by integrating over the
systematic parameters in the model. 
 
The resulting samples then enable us to study single- 
and multi-dimensional marginal distributions of the parameters of interest
and extract all kinds of statistical estimators like mean values, credibility 
intervals, and so forth.

The posterior parameter distribution is obtained from Bayes' theorem, 
\begin{equation}
\label{Eq:Bayes}
p(\theta|D,M) = \frac{p(D|\theta,M)\ p(\theta|M)}{p(D|M)} \; .
\end{equation}
Here $p(\theta|M)$ is the prior distribution of the parameters~$\theta$
and $p(D|\theta,M)$ is the
probability to observe the data $D$, given a particular set of parameter values $\theta$.  
In \Eq{Eq:Bayes} it is evaluated as a function of $\theta$ 
for fixed data $D$ and is called the likelihood function (it is not a probability
density), denoted as
\[
\mathcal{L}_D(\theta) = p(D|\theta), \quad \text{$D$ fixed}.
\]
Here and in the following, we will always imply the model choice $M$
but for clarity, it will be dropped in the notation.

An important feature of Bayes' theorem is the fact that the support of the
posterior can never exceed that of the prior. This makes it straightforward to impose
hard constraints on allowed parameter values, such as the requirement
that a parameter is positive definite.

The normalization of the posterior is given by the so-called evidence
\begin{equation}
  p(D) = \int \mathcal{L}_D(\theta)\ p(\theta)\ \der \theta \; .
\end{equation}
The evidence is a scalar value that is often challenging to compute as it results from a
multi-dimensional integral. However, as it is independent of the parameter vector $\theta$, its
value is not needed to study the relative credibility of parameter values. Sampling
algorithms like Markov Chain Monte Carlo (MCMC) make it possible to draw samples distributed
according to the posterior distribution based on the non-normalized posterior density
\begin{equation}
\tilde{p}(\theta|D) = \mathcal{L}_D(\theta)\ p(\theta)
\end{equation}
This yields posterior samples that can be used to produce the desired marginal parameter distributions, which can always be normalized afterward. 
%
%-------------------------------------------------------------------------------
%
\subsection{Implementation of MCMC \label{se:mcmc}}
In our experience, prior parameter distributions with 
hard constraints, like positive definite parameters that may assume values
close to zero can negatively impact MCMC convergence and efficiency. 
Several of the parameters of our model fall into this category. 
Furthermore, the momentum sum rule (see below) introduces a strict correlation between the momentum parameters that restricts the support of the posterior density to a subspace of the full parameter space. 
Densities with such a support typically cause sampling algorithms to fail completely, as the posterior density is zero almost everywhere in the parameter space. The methods to circumvent this obstacle are described below.

We solve the problem of sampling from a complicated distribution by performing a change of variables with the aim of creating a parameter space that is straightforward to sample from.
Thus, instead of sampling the density $\tilde{p}(\theta|D)$ directly, 
%
% I like to get rid of the subscript 'pt' in p_pt and f_pt
% To put it back just set the argument in \ppt{pt} and \fpt(pt)
%
we introduce a suitable transformation $\theta = \fpt{}(x)$ and sample
the density
\begin{equation}
\ppt{}(x|D) = \mathcal{L}_D[\fpt{}(x)]\;\mathcal{N}(x)
\end{equation}
where $\mathcal{N}(x)$ is a multivariate Normal distribution of the
appropriate dimension.  
The posterior $\ppt{}(x|D)$ is unbounded and has full support over
it's parameter space, and so is much easier to sample. Samples of the original
parameters $\theta$ are then obtained by simply applying the parameter
transformation $\fpt{}$ to generated samples of the parameters $x$ in the 
alternate space.

The challenge lies in finding suitable transformations $\fpt{}$.
For univariate components of the prior
distribution, their quantile functions and the cumulative distribution function of the normal distribution 
provide the necessary building blocks~\cite{Schulz:2021BAT}. For the Dirichlet 
distribution that we use to satisfy the momentum sum rule in the prior (see
\secref{sec:forward_model}), a suitable transformation is given 
in~\cite{ref:Betancourt}. 

\subsection{Marginalization and uncertainty propagation}
\label{sec:Marginalization}

In many model analyses
one is interested not in the full posterior distribution but in the 
marginal probability
distribution of only one or a few parameters.
For example, the probability distribution of a single
parameter $\theta_i$ is obtained from 
\[
  p(\theta_i|D) = \int p(\theta|D) \der {\theta}_{j\neq i} \, .
\]  
In practice, this calculation is performed quite simply by histogramming 
the $\theta_i$  values from all posterior samples, ignoring the other
parameters. From such a
histogram, several estimators can be computed to report the results.  
Commonly used quantities are:
\vspace{0.5em}
\paragraph{Mode of $\theta_i$.} 
The value ${\theta}_i^*$ where the marginalized posterior probability
density has a maximum. Modes are usually computed for fully marginalized
posteriors $p(\theta_i|D)$ (`marginal mode'), for the entire posterior
(`global mode'), or for any number of parameters, with the rest marginalized.
Note that the parameter
values that maximize the full posterior distribution usually do not coincide
with those that maximize marginalized distributions.
\vspace{0.5em}
\paragraph{Mean of $\theta_i$.} 
This is the expectation value
\[
  <\theta_i> = \int_{\tht{min}}^{\tht{max}} p(\theta_i|D) \theta_i 
  \der \theta_{i}\, ,
\]
with the  parameter bounds denoted by
$[\theta_{\mathrm{min}}, \theta_{\mathrm{max}}]$.%
\vspace{0.5em}
\paragraph{Median of $\theta_i$.}
The value $\hat{\theta}_i$ that splits the probability content
of $p(\theta_i|D)$ in two:
\[
  \int_{\tht{min}}^{\hat{\theta}_i} p(\theta_i|D) \der \theta_{i} = 
  \int_{\hat{\theta}_i}^{\tht{max}} p(\theta_i|D) \der \theta_{i} = 0.5\,.
\]
\vspace{0.5em}
\paragraph{Central Interval.}
The $(1-2\alpha)$ central interval $[\tht{-},\tht{+}]$
is defined such that a fraction $\alpha$ of the probability is contained
on either side of the interval:
\[
 \int_{\tht{min}}^{\tht{-}} p(\theta_i|D) \der \theta_{i}=
 \int_{\tht{+}}^{\tht{max}} p(\theta_i|D) \der \theta_{i} = 
 \alpha.
\]
\vspace{0.5em}
\paragraph{Smallest Interval.}
The $\alpha$ smallest interval(s) is defined such that a fraction $\alpha$ of
the probability is contained in a set of intervals where the set size is
minimized.  This is realized as a Lebesgue integral:
\[
\int_{p(\theta_i|D)\geq p_{\rm min}} p(\theta_i|D) \der \theta_{i}= \alpha
\]
where $p_{\rm min}$ is to be determined.  This procedure can result in several intervals
in $\theta_i$ in the case of multimodal distributions.
\vspace{0.5em}
\paragraph{Uncertainty Propagation.}
Having full access to the posterior allows for the 
evaluation of the probability distribution of any function of the parameters.
In contrast to standard techniques used for error propagation, there is no
need here for any assumptions like distributions being Gaussian-shaped.
As an example, consider that we have a function $f(x| \theta)$ of interest,
for instance, a parton distribution that depends on a subset of the $\theta$
parameters. The distribution of $f(x)$ is
then given by
\[\
p(f(x)|D) = \int p(f(x|\theta)) p(\theta|D) \der \theta
\]
which can be evaluated in a straightforward way from the posterior samples.
We give examples of such uses below.
%
%-------------------------------------------------------------------------------
%
\subsection{BAT.jl - the Bayesian Analysis Toolkit}
\label{sec:BAT}

For the Bayesian inference process, we use the software package
\batjl~\cite{Schulz:2021BAT}, which is a high-performance toolkit for
Bayesian analysis tasks, coded in the \julia\ programming
language~\cite{ref:Julia}. The package
provides multiple algorithms for posterior sampling, integration and 
mode-finding, as well as automatic plotting and reporting functionality.

The \batjl\ package also has the ability to automatically generate suitable space
transformations between user-defined probability distributions and standard
multivariate normal or uniform distributions. This enables us to automatically
perform prior-space transformations as described in \secref{se:mcmc} and sample the
posterior via the Metropolis-Hastings algorithm in an unconstrained space where
the prior has become
a normal distribution in all dimensions. The samples are then automatically
transformed back into the original space.

A full description of the algorithms and tuning of the \batjl\ toolkit
can be found in~\cite{Schulz:2021BAT}. 

%
%-----------------------------------------------------------------------------
%
\section{Analysis Structure\label{se:epAnalysis}}

As mentioned in the introduction, we focus in this initial release of our \pdf
determination code on the analysis of electron(positron)-proton 
deep inelastic scattering data. The data are generally reported in bins of the scaling variables $x$ and $Q^2$, which are computed from the reconstructed four momenta of the incoming proton and the incoming and scattered leptons (for more details see, e.g.,~\cite{ZEUS:2020ddd}). 
In this section, we describe how the event 
numbers observed in these data are predicted, starting from a
set of proton \pdf[s] parameterized at some input scale $Q^2_0$.
%
%------------------------------------------------------------------------------
%
\subsection{Computation of the \epmp\ cross sections}
\label{sec:pdfs_and_evo}

To compute the \epmp\ deep inelastic cross sections,
we start from a set of quark, antiquark, and gluon distributions 
$xf_i(x)$, parameterized at a fixed 
value of $Q^2_0$. The aim of the analysis is to determine from the data
the posterior joint distribution of these parameters.

In the above, $f_i(x)$ is the number density of partons
of type $i$ in the proton, and $xf_i(x)$ is the 
momentum density of these partons.

The next step is to evolve these distributions in perturbative 
QCD~\cite{Gribov:1972ri,Gribov:1972rt,Lipatov:1974qm,
Dokshitzer:1977sg,Altarelli:1977zs} 
from the input scale to larger values of~$Q^2$. We use the
\qcdnum\ program~\cite{Botje:2010ay} to evolve the \pdf[s]
in the \msbar-scheme at next-to-next-to-leading 
order (NNLO)~\cite{Tarasov:1980au,Curci:1980uw,
Furmanski:1980cm,Furmanski:1981cw,Larin:1993tp,Chetyrkin:1997sg,Buza:1996wv,
Moch:2004pa,Vogt:2004mw,Ball:2015tna}.

In the analysis, we impose the momentum
sum rule and the valence quark
counting rules. The momentum sum rule states that the fractional 
momenta of all partons in the proton add up to unity:
\begin{equation}
\label{eq:mom_sum_rule}
   \sum_i \int_0^1  xf_i(x) \der x =  \sum_i \Delta_i = 1 \; .
\end{equation}
We introduce here the notation $\Delta_i$ for the total momentum fraction
carried by the parton species~$i$.

The quark counting rules fix the net number of quarks in the proton so
that its quantum numbers are conserved:
\begin{equation}\label{eq:vsum}
  \int_0^1 [ q_i(x) - \bar{q}_i(x) ]\; \der x = \left\{
  \begin{array}{l}
    2\ \ \text{for}\ \ i = u\; , \\
    1\ \ \text{for}\ \ i = d\; ,\\
    0\ \ \text{for}\ \ i = s,c,b, t\; .
  \end{array}
  \right.
\end{equation}
Here and in the following we use the notation $q$, $\bar{q}$ and $g$
to denote quark, antiquark, and gluon densities.

It is important to point out that the QCD evolution equations
guarantee that sum rules that are imposed
at the starting scale $Q^2_0$ will be satisfied at
all scales. 

The neutral current DIS cross section for \epmp\ scattering is given in terms
of generalized structure functions of the proton as ($y$ is the inelasticity variable, see~\cite{ZEUS:2020ddd})
\begin{equation}\label{eq:nc}
\frac{{\der}^2\sigma^{\epmp}_{\mathrm{NC}}}{\der x \der Q^2} = 
\frac{2\pi\alpha}{xQ^4}(Y_{+}\tilde{F}_2 \mp Y_{-}x\tilde{F}_3 - 
      y^2\tilde{F}_{\mathrm{L}})   
\end{equation}
where $Y_{\pm}=1\pm (1-y)^2$, and $\alpha$ is the fine-structure constant.

The generalized structure functions are related to the vector and  
axial-vector coupling constants \vve\ and \aae\ by
\begin{equation}   \label{eq:StrucFunc}
\begin{split}
 \tilde{F}_i &= \Fg{i} -\kaZ \vve\, \FgZ{i} + \kaZ^2 (\vve^2 + \aae^2)\, \FZ{i},
 \quad i = 2,\mathrm{L} \\
 x\tilde{F}_3 &= \kaZ \aae\, x\FgZ{3} + \kaZ^2 2\vve \aae\, x\FZ{3} \mbstrut{1.0em}
\end{split}
\end{equation}
where
\[
\kaZ(Q^2) = \frac{Q^2}{(Q^2+\emz^2)\, 4\sin^2{\tht{w}}\cos^2{\tht{w}}}.
\]
In our analysis, we use 
$\sin^2\tht{w}= 0.231$ and $\emz=91.1876$~GeV
for the electroweak mixing angle and the Z-boson mass, 
respectively~\cite{ParticleDataGroup:2022ssz}. 

In LO QCD, the structure functions 
are linear combinations of parton densities of different
flavor (note that $\eFL = 0$ at LO).
\begin{equation}\label{eq:StrucFunc2}
 \begin{split}
 \{F_2^\gamma,F_2^{\gammaZ},F_2^Z\} &= x \sum_i 
 \{e_i^2,2e_i \ggV, \ggA^2\}(q_i+\overline{q}_i)\\
 x\{F_3^\gamma,F_3^{\gammaZ},F_3^Z\} &= x \sum_i 
 \{0,2e_ig_A,2\ggV \ggA\}(q_i-\overline{q}_i)
 \end{split}
\end{equation}
where $e_i$ is the charge of the quark species~$i$
and \ggV\ and~\ggA\ are the weak vector and axial-vector
couplings of the quark to the Z-boson. 

We use the \qcdnum\ package to compute the structure functions at NNLO
which involves convolutions of parton densities with 
various coefficient functions~\cite{F123:1,
F123:2,F123:3,F123:4,F123:5,F123:6}.
Note in this respect that the coefficients
in \Eq{eq:StrucFunc2} are the same for all up-type (down-type) quarks with
$e_i = \frac{2}{3}\ (\frac{1}{3})$. Thus we can compute NNLO structure functions
separately for the sum of up-type and down-type quarks and multiply these by
the coefficients afterwards. This allows for an efficient calculation of the
cross section based on the NNLO computation of only 6 structure functions.
%
%------------------------------------------------------------------------------
%
\subsection{Forward model}
\label{sec:forward_model}

In the forward modeling approach, we compute the
expected number of events $\nu$ in a bin 
$(\Delta x,\Delta Q^2)$ as an integral of the differential
cross section over the 
full kinematic phase space $\mathcal{D}$. Introducing the short-hand
notation $\ve{u} = \{x,Q^2\}$ this integral can be written as
\begin{equation}\label{eqn:nu_integ}
 \nu =\mathcal{L} \int_{\Delta \ve{u}'} \left[
 \int_{\mathcal{D}} A(\ve{u}'|\ve{u}) 
 \frac{\der^2 \sigma(\ve{u})}{\der \ve{u}}  
 \der \ve{u} \right]   \der \ve{u'}\, ,
\end{equation}
where the primed (unprimed) variable refers to the reconstructed (true)
kinematics.
Here $\mathcal{L}$ is the integrated luminosity of the data set being
analyzed, and $A$ is a transformation kernel that takes into account the detector
effects and radiative corrections to the QED Born-level differential cross 
sections, computed as described in \secref{sec:pdfs_and_evo}. The forward model described here can be implemented for other experiments or independently of \partondensity\, as long as the transfer matrix $A$ is provided by the experimental groups as part of their data release.

We assume here that this information is cast in the form of a matrix $A$ that
provides a mapping from the Born-level cross section integrated over bins
in the kinematic domain to events accumulated in the experimental bins.
Denoting by~$\nu_i$ the counts in an experimental bin~$i$ and by~$\lambda_j$
the integrated cross section in a kinematic bin~$j$,
\Eq{eqn:nu_integ} is written as
\begin{equation}\label{eqn:nu_approx}
 \nu_i = \mathcal{L} \sum_j A_{ij} \lambda_j 
\end{equation}
Experimental systematic uncertainties are encoded in a set of deviation
matrices. The total systematic deviation is then
given by a linear combination of these
deviations with weights that are included
in the set of model parameters $\theta$.
In \secref{sec:ZEUSexample}, we will describe in more detail how the forward modeling
is implemented in our present analysis framework.

Because we are analyzing event counts, it is trivial to write down the
likelihood as a product of Poisson distributions:
\begin{equation}\label{eqn:Plikelihood}
   \mathcal{L}_D(\theta) = \prod_{\text{bins}} 
      \frac{\nu^{n} e^{-\nu}}{n!}
\end{equation}
where $n$ is the number of events observed in a bin and $\nu$ is that
predicted by the forward model, using a particular set of model parameter
values $\theta$. 

Analyzing binned event counts has the advantage that the data points
are uncorrelated and have known probability distributions and that there are no problems with including sparsely populated
or empty bins. To our knowledge, these features are currently unique
to our approach. 
%
%-----------------------------------------------------------------------------
%
\section{Technical Developments\label{se:technical}}

We now describe the main numerical and programming steps carried out while developing our analysis code.

\subsection{The SPLINT package }

Integrating the cross sections with standard numerical methods like two-dimensional
Gauss quadrature turns out to be quite CPU-time-consuming because a sizable 
sample of NNLO structure functions has to be computed for each integration.
To solve this problem, we added to the \qcdnum distribution
the \splint package to construct cubic
interpolation splines of structure functions and cross sections. Sampling
from splines is much faster than \xit{ab-initio} computation, while
\splint\ provides fast routines for cubic spline integration.

A cubic interpolation spline $S(u)$ in the variable $u$
is a piecewise cubic polynomial defined on a strictly ascending set of
node-points $\{u_i\}$. The four polynomial coefficients in each node-bin are
adjusted such that the spline coincides at each node $u_i$ with an input
value $f_i$, and is continuous in the first and second
derivative; the third derivative is allowed to be discontinuous
at the node points. These conditions
lead to a set of linear equations in the coefficients, which can be solved if
boundary conditions are given on the slopes at the two end points of the spline. 
We use a spline algorithm that fixes the third derivatives to values estimated
from divided differences.
A one-dimensional spline $S(u)$ becomes a two-dimensional spline
$S(u,v)$ by parameterizing the $u$-coefficients 
as cubic splines in~$v$.

The \splint\ package constructs splines on a
selected set of \qcdnum $x$-$Q^2$ evolution grid-points. 
Alignment of the nodes to the evolution grid avoids
\qcdnum interpolation of a 
user-given input function $f(x,Q^2)$ when constructing the spline. A coarser
node-grid gives a faster spline construction
but also a less precise spline approximation 
so that some tuning is necessary to balance speed versus accuracy,
see \secref{se:tune}.

Spline interpolation becomes interesting when many samples are needed of functions
that are expensive to compute, such as structure functions and cross sections.
For this
purpose \splint has a special routine for structure function input that makes
use of the very fast list-processing capabilities of \qcdnum. Apart from creating
splines, the \splint package also has routines
to integrate these over rectangular bins taking, if necessary, into account
the kinematic limit $Q^2 \leq xs$, where $\sqrt{s}$ is the centre-of-mass energy
of the \epmp\ collisions.

Inside \splint, the splines are functions of the internal \qcdnum variables
$u = -\ln x$ and $t = \ln Q^2$, which introduces a Jacobian $e^{-u} e^t$ in
the integrals. As a consequence, spline integration is expressed in terms
of the fundamental ($\gamma$-function) integrals
\begin{equation}\label{eq:Epm}
  E^{\pm}(z,n) = \int_0^z w^n e^{\pm w}\, \der w.
\end{equation}
Partial integration of \Eq{eq:Epm} yields simple recursion relations 
between the $E^{\pm}$ at successive
values of $n$ so that they can be computed rapidly for all
powers 0--3 needed for the term-by-term integration of a cubic polynomial inside
a node-bin. When a node-bin crosses the kinematic limit, we have to integrate both
over rectangles and right-angled triangles in the $u$-$t$ plane.
The triangular domains are handled by \splint integration over $u$ and Gauss
integration over $t$.
  
To validate the procedure, we integrated splines over arbitrary rectangles, with
and without crossing the kinematic limit, both with \splint and with a
two-dimensional Gauss integration routine. In all cases, the relative difference
was found to be $< 10^{-9}$ with \splint running about a factor of~300
faster than Gauss. 

For a more detailed description of the integration algorithm in \splint 
we refer to the write-up, which is available from the \qcdnum website.\footnote{
  \texttt{https://www.nikhef.nl/user/h24/qcdnum}
  }   
%
%----------------------------------------------------------------------------
%
\subsection{The tuning of QCDNUM and SPLINT\label{se:tune}}

In the range $x > 10^{-3}$ and $100 < Q^2 < 3\times10^4\ \GeV^2$
we have tuned the \qcdnum evolution grid and the \splint spline nodes with
the aim to obtain a relative accuracy on the cross sections of better than
$5\times10^{-4}$, with the code still running at an acceptable speed.
For this tuning, the cross sections are computed at $\sqrt s = 300\ \GeV$.

In~\cite{Botje:2010ay} it is recommended to run the \qcdnum\ evolution
on a $100 \times 50$ $x$-$Q^2$ grid with break points at 
$x = \{ 0.2,0.4,0.6,0.8 \}$ where the grid density doubles towards larger $x$
at each break point. Comparison with evolution on a $300\times150$ grid shows
a relative \pdf accuracy of better than $10^{-4}$, 
except at very large $x$ where the parton distributions vanish.

\begin{figure}[htb] 
    \centering
    \includegraphics[width=0.475\textwidth]{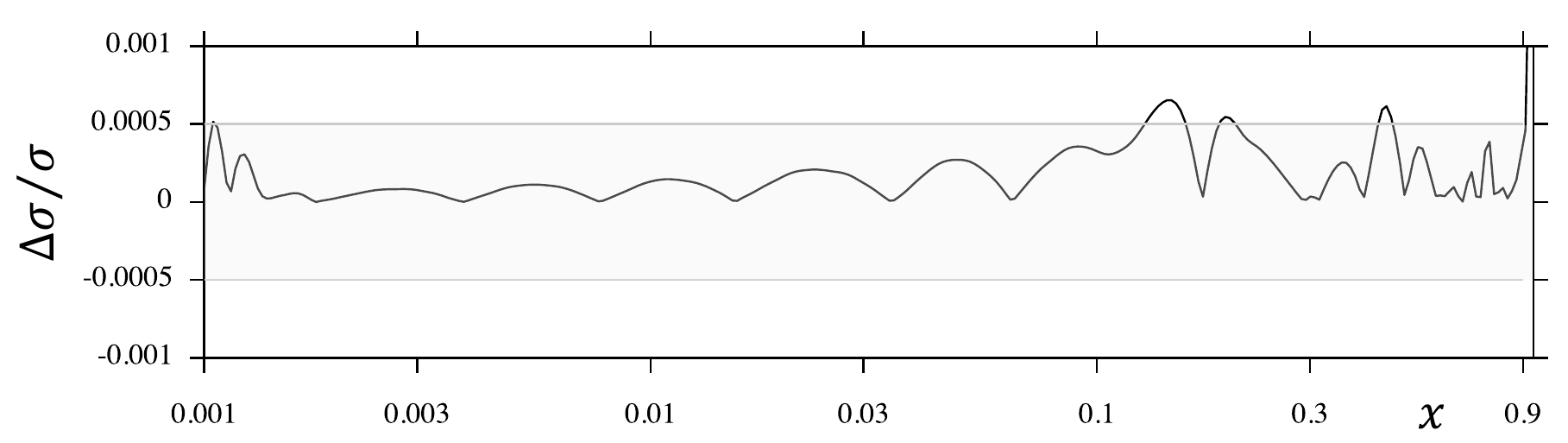}
    \caption{Estimate of the relative error on the differential cross section
             versus $x$ at $Q^2 = 100$~GeV$^2$. Above $x = 0.95$ the cross section
             vanishes, and a relative error becomes ill-defined.}
    \label{fig:dxsec}
\end{figure}

Because structure functions are slowly varying in~$x$ and~$Q^2$, 
a coarse $22\times7$ node grid is sufficient to yield interpolation splines
with a relative accuracy
better than $5\times10^{-4}$, and the same is true for the cross section spline on
a $100 \times 25$ node grid, see~\fig{fig:dxsec}. The latter grid must be dense since the cross section is strongly varying.
%-------------
%-------------

In \tabref{tab:xsecCPU}, we show the CPU time needed when running
the tuned code on a 2018  MacBook Pro with an Intel processor. Starting from a set of input \pdf[s], computing 429 integrated cross sections takes less than 10~ms on such a machine. 
\begin{table}[tbh]
\caption{CPU cost of computing integrated cross sections. }
\begin{ruledtabular}
\begin{tabular}{lcc}
\xbf{Subtask}   & \textbf{Grid} & \xbf{CPU time [ms]}  \\
 \hline
Evolution & $100 \times 50$ & 3.6 \\
Structure function splines (6$\times$) & $22 \times 7$ & 2.9 \\
Cross section spline & $100 \times 25$ & 2.2 \\
Integration over 429 bins & & 0.8
\end{tabular}
\end{ruledtabular}
\label{tab:xsecCPU}
\end{table}%
%
%----------------------------------------------------------------------------
%
\subsection{The QCDNUM.jl wrapper}
We have developed the \qcdnumjl package\footnote{
  \url{https://github.com/cescalara/QCDNUM.jl}
} 
which is the
\julia interface  to the \qcdnum fast QCD evolution and convolution routines.
The  \qcdnum\ program is written in \fortranzz and the
interface gives us access to this fast, versatile and 
well-tested software in the \bat analysis framework (see \secref{sec:BAT}).

With \qcdnumjl, we provide a lightweight but easily usable wrapper, 
with all the original \qcdnum functionality available to a
\julia programmer. 
Example programs are provided in the documentation, 
allowing those familiar with \qcdnum to quickly adapt their code. As part of the \julia interface, we also provide a well-documented,
high-level interface to implementing and visualizing the evolution of \pdf[s]
that may be appealing to both new and old users of \qcdnum. 

The implementation in \julia\ offers several further advantages. 
Thanks to \binarybuilderjl\footnote{
  \url{https://binarybuilder.org}
}, 
\qcdnum is automatically compiled and installed in a cross-platform manner,
with no actions needed from the user. It is also possible to interface code
relying on \qcdnum with \mbox{\julia's} rich functionality and ever-growing
modern package ecosystem. The existing tools for working with \julia\ code,
such as Jupyter notebooks, enable more accessible and
reproducible code. As such, the \qcdnumjl package is a useful
addition to the community bringing the use of \qcdnum to a
broader audience and modern coding practices. 

\subsection{PartonDensity.jl package \label{se:PDpackage}}

We implement a full forward model and interface to \bat in
\partondensity\footnote{
  \url{https://github.com/cescalara/PartonDensity.jl}
} 
to enable the simulation of data and inference of \pdf parameters.
Included are the interfaces to \qcdnum and \splint, as well as
the different \pdf parameterizations described in this work.
We also implement the forward modeling transfer matrices from the ZEUS
experiment, described in \secref{sec:ZEUSexample}.
Within this framework, we provide prior and likelihood definitions that can be
used with \bat, as well as tools for visualization. The software is well
documented and designed in a modular way to allow extension to other \pdf
parameterizations and data sets of interest.  

\begin{figure*}[htb] 
    \centering
    \includegraphics[width=0.7\textwidth]{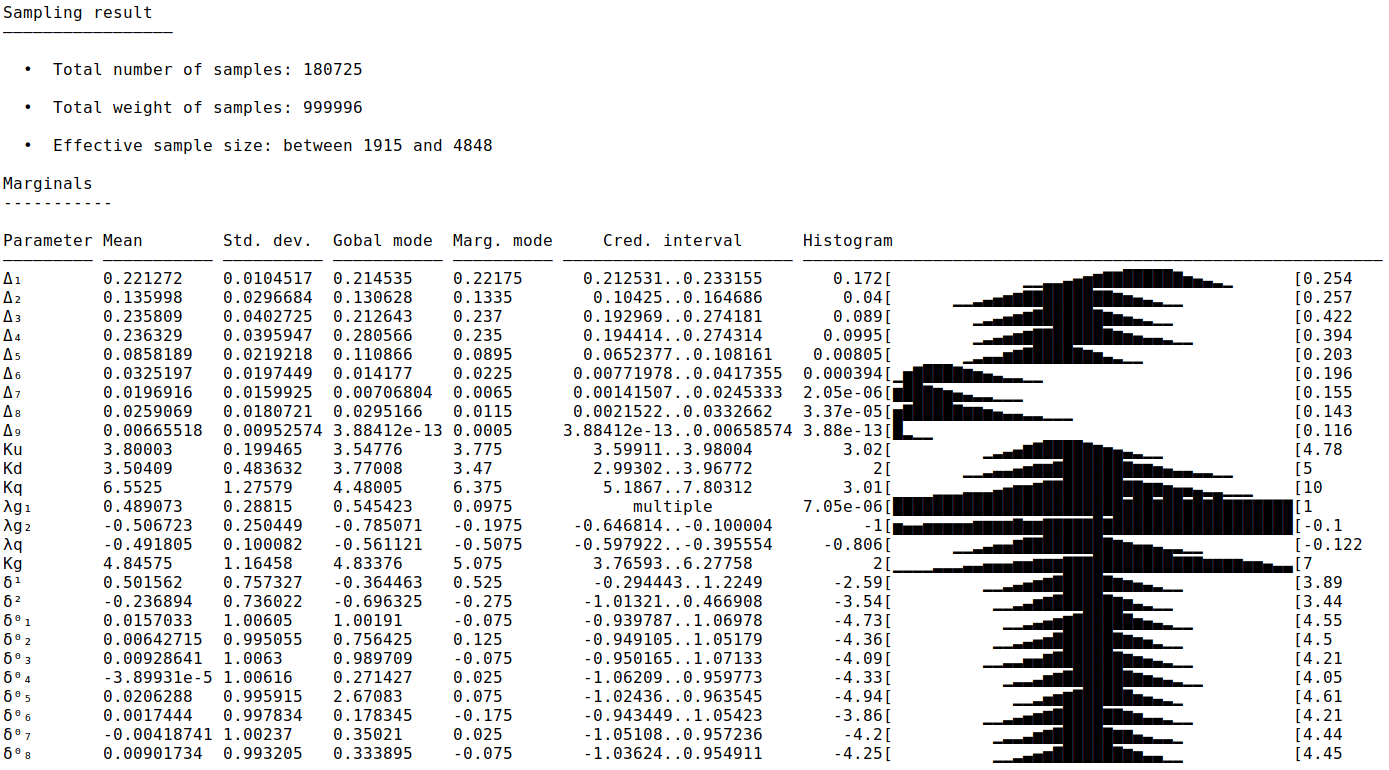} 
    
    \vspace{5mm}
    
    \includegraphics[width=0.75\textwidth]{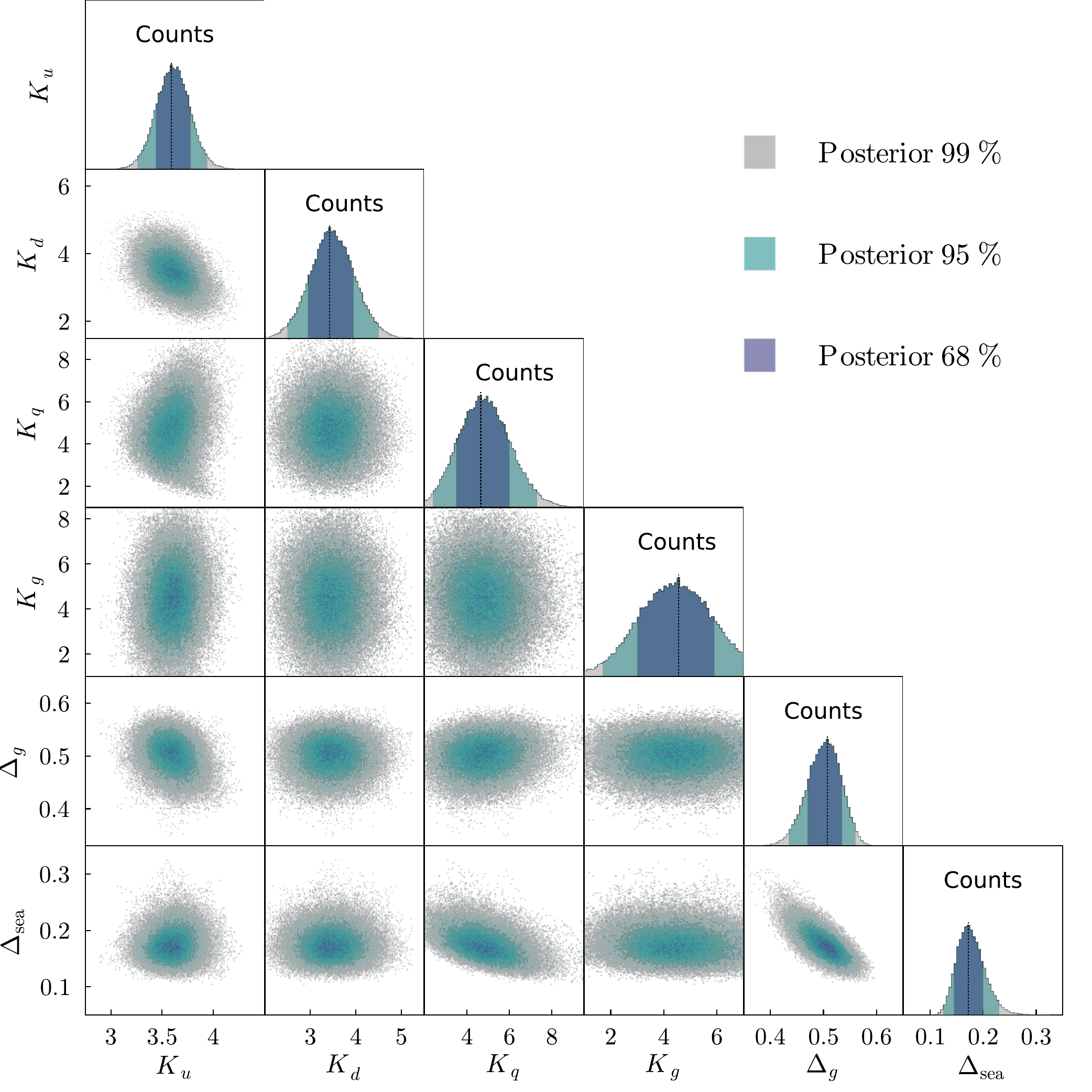}
    \caption{Upper panel: Example of a statistical summary of a 26-dimensional posterior
             distribution. Lower panel: Parameter correlation distributions for a subset of the model parameters. The marginal distributions are shown on the diagonal. Taken from the pseudo-data analysis
             described in \secref{se:validation}.}
    \label{fig:reportcorner}
\end{figure*}

Several forms of standard output are available from our code.  
One useful feature is a text-file summary of the posterior distribution, an example
of which is shown in the upper panel of~\fig{fig:reportcorner}.
%---------------------
%----------------------
The marginalized posteriors and global mode values are given for all parameters.  
Many graphical representations of the output are available in standard form, 
and others can be easily generated.  
As an example, we show in the lower panel of~\fig{fig:reportcorner} a two-by-two correlation plot that can  
be made for any subset of the parameters.

%-------------------------------------------------------------------------
%
\section{Parton Density parameterizations\label{sec:PDFs}}

The \pdf determination procedure requires a well-defined set of parameterizations
for the different parton densities at the input scale $Q^2_0$ of
the evolution.  A variety of forms are currently used
by the different fitting
teams~\cite{Cridge:2021pxm,Bailey:2020ooq,Hou:2019efy,Dulat:2015mca,
Alekhin:2018pai,Alekhin:2013nda,NNPDF:2021njg,NNPDF:2017mvq,
H1:2021xxi,H1:2015ubc} which are all of the type:  
\begin{equation}
    xf_i(x) = A_i x^{\lambda_i}F_i(x)(1-x)^{K_i}. 
\end{equation}
Each parton species $i$ has its
own set of parameters and function $F$. The behavior of a \pdf as
$x \rightarrow 1$ is largely controlled by its parameter $K$, 
while $\lambda$ controls the behavior as $x \rightarrow 0$.
The function $F(x)$ interpolates between these two extreme regions,
and the  parameter $A$ fixes the normalization.

In \secref{sec:pdfs_and_evo} it is mentioned that the
quark and gluon distributions $xf_i(x)$ are required to satisfy the
momentum sum rule and quark counting rules as given in 
\Eqs{eq:mom_sum_rule}{eq:vsum}, respectively. 
We therefore wish to construct parameterizations that are flexible, 
positive-definite, easily evaluated, and quickly integrated.
We also wish, at this stage in the development of the Bayesian framework, to
restrict the number of \pdf parameters as much as possible. Given these requirements,
we have fully implemented in \partondensity a Beta-distribution parameterization
that provides a suitable starting point in a future exploration
of a variety of alternatives. 

%-------------------------------------------------------------------------------

\subsection{Parameterization based on Beta distributions\label{sec:BetaPDFs}}
To parameterize the quark densities, it is convenient to write them
as valence (\Sup{q}{v}) and sea ($\Sup{q}{s}$) distributions,
\[
  q + \bar{q} = (q - \bar{q}) + 2\bar{q} = \Sup{q}{v} + \Sup{q}{s}.
\] 

We parameterize the valence momentum densities as
\begin{equation}\label{eq:xvalence}
  x\Sup{q}{v}_i(x,Q^2_0) = \left\{ \begin{array}{ll}
     A_i \;x^{\lambda_i} (1-x)^{K_i} & \ \text{for} \ i = u,d \\
     0                               & \ \text{otherwise}.
     \end{array}\right.
\end{equation}
The integral of the number density, $\Sup{q}{v}_i(x,Q^2_0)$ is finite for
$\lambda_i > 0$. Similarly, for the antiquark distributions, we have
\begin{equation}
    x\bar{q}_i(x,Q^2_0) = A_i\; x^{\lambda_{\bar{q}}} (1-x)^{K_{\bar{q}}}
    \quad \text{for} \quad i = \bar{u},\bar{d},\bar{s},\bar{c},\bar{b},
    \label{eq:antiquark_dist}
\end{equation}
where all antiquark flavors share the same $x$-dependence, but have different
normalisations $A_i$. 
For this parameterization, we require $-1 < \lambda_{\bar{q}} <0$ so that
$x\bar{q}$ is integrable and increasing at low $x$. Finally, we parameterize the
gluon density as the sum of valence and sea contributions
\begin{equation}
\begin{split}
    xg(x,Q^2_0) &= x\Sup{g}{v}(x) + x\Sup{g}{s}(x) \\
    &= \Sup{A}{v}_g\; x^{\Sup{\lambda}{v}_g}(1-x)^{K_{g}}
    + \Sup{A}{s}_g\; x^{\Sup{\lambda}{s}_g}(1-x)^{K_{\bar{q}}}.\quad
\end{split}    
\end{equation}
where we set 
$-1 < \Sup{\lambda}{s}_g < \Sup{\lambda}{v}_g$ and set further restrictions on these 
parameters in order to obtain an
integrable gluon density with a valence (sea) contribution that decreases
(increases) towards low $x$. Here $K_g$ is an independent parameter, 
while $K_{\bar{q}}$
is shared with the $x\bar{q}_i$ densities defined in \Eq{eq:antiquark_dist}.

For a Beta-distribution $xf(x) = A\, x^{\lambda}(1-x)^K$ we can replace
the normalization constant by the momentum integral through the relation
\begin{equation}\label{eq:DeltaFromA}
    \Delta = A\, \frac{\Gamma(\lambda+1) \Gamma(K+1)}{\Gamma(\lambda+K+2)}.
\end{equation}
The valence sum rule, introduced in \Eq{eq:vsum}, 
relates the normalizations to the shape parameters by 
\begin{equation}\label{eq:Aupdown}
A_i = \Sup{N}{v}_i\; \frac{\Gamma(\lambda_i+K_i+1)}
{\Gamma(\lambda_i)\Gamma(K_i+1)}
\qquad i = u,d\;,
\end{equation}
with $\Sup{N}{v}_u = 2$ and $\Sup{N}{v}_d = 1$.

Using the property
$\Gamma(z+1) = z\Gamma(z)$ we find 
from \Eqs{eq:DeltaFromA}{eq:Aupdown} the following relation between
the total momentum $\Delta$ carried by the $u$ or $d$ valence quarks
and the shape parameters
\begin{equation}\label{eq:getlambda} 
\Delta_i =  
\Sup{N}{v}_i\; \frac{\lambda_i}{\lambda_i + K_i + 1} \qquad i =u,d.
\end{equation}
%
%-------------------------------------------------------------------------------

\subsection{Prior parameter constraints}\label{sec:priorconstraints}

Given a set of free \pdf parameters, the challenge is to include as many physically motivated constraints on the priors as possible.

For this, it is advantageous to replace the normalization constants, $A$, which are not straightforward for interpretation, with the momentum fractions, $\Delta$,
either by numerically integrating the \pdf[s] or by using
\Eq{eq:DeltaFromA} in case the \pdf[s] are parameterized in terms of 
Beta-distributions. 

In that case \Eq{eq:getlambda} offers two alternatives
for the parameters of the up and down valence distributions:
(i)~leave the shape parameters $\lambda$ and $K$ free and thereby fix the
momentum fraction $\Delta$ or (ii)~leave $\Delta$ free and thereby fix
one of the two shape parameters. 

The first alternative allows to specify
the shape of the valence densities, but it is not trivial to ensure that
$\Delta_d + \Delta_u < 1$ and $\Delta_d < \Delta_u$ over the full support
of the prior.

In the analysis of the ZEUS high-$x$ data, we have chosen to fix the low-$x$ shape
parameter $\lambda$ and leave free the \mbox{high-$x$} shape parameter $K$
and the momentum fraction~$\Delta$. In this way we
include in the model parameters the complete set of \pdf momenta, 
subject to the sum rule
constraint of~\Eq{eq:mom_sum_rule} given in \secref{sec:pdfs_and_evo}.

It is convenient to use a Dirichlet distribution as a joint prior for the momenta~\cite{ref:Betancourt}. 
A Dirichlet distribution ${\rm Dir}(\alpha_1,\ldots,\alpha_k)$ of 
$k$ independent variables $u_i \in [0,1]$  
lives on a $(k-1)$-dimensional manifold defined by
$\sum u_i = 1$ so that
the momentum sum rule is automatically satisfied. It is a multivariate
generalization of the Beta distribution; 
for instance ${\rm Beta}(\alpha_1,\alpha_2)$ of one
variable $u$ is the same as ${\rm  Dir}(\alpha_1,\alpha_2)$ of two variables $(u_1,u_2)$ with $u_1 + u_2 = 1$.

With a suitable choice of the shape parameters $\alpha$, it is possible 
to satisfy expectations such as that,
asymptotically, gluons and quarks carry approximately
the same momentum, that valence up-quarks carry about twice
the momentum of valence down-quarks, and that the heavier quarks carry
little momentum.

The spectator counting rules of
Brodsky and Farrar~\cite{BF} give a first expectation for the ranges of the shape parameters
$K_u, K_d, K_{\bar{q}}$ and $K_g$. Furthermore, a body of
PDF results available from the literature indicate the preferred range for these parameters.

%-----------------------------------------------------------------------------
\section{Validation \label{se:validation}}

To validate our Bayesian tools, we analyzed sets of simulated data
and compared the posterior distributions to the
parameter input values. Because the tools were initially developed for
the analysis~\cite{Aggarwal:2022cki} of the ZEUS high-$x$ and
large-$Q^2$ data~\cite{ZEUS:2013szj},
our simulations were restricted to the kinematic range 
covered by these data. An adequate representation is obtained by
using the simple Beta parameterization described in the previous sections,
at an input scale of $Q^2_0 = 100$~GeV$^2$.
Some details of the ZEUS experiment are given below.

%-----------------------------------------------------------------------------------------------
\subsection{The ZEUS experiment and systematics}
\label{sec:ZEUSexample}

In~\cite{ZEUS:2013szj}, the ZEUS collaboration has published
\epmp\ deep-inelastic scattering data covering
the range $0.03 \leq x \leq 1$ and $650 \leq Q^2 \leq 20000$~GeV$^2$.
These data are unique in providing measurements up to $x=1$
in the high $Q^2$ regime where higher-twist effects
are absent so that a clean analysis can be performed based on
the perturbative QCD evolution equations.  
In~\cite{Aggarwal:2022cki}, it is shown that the data mainly constrain
the parameters of the valence up-quark, as will also become apparent in the
figures and results given below.

The data are presented as counts in 153 bins,
separately for the e$^-$p and e$^+$p data sets.

To enable forward modeling, ZEUS has provided in~\cite{ZEUS:2020ddd}
the transfer matrix $A$ as defined
by \Eq{eqn:nu_approx} in \secref{sec:forward_model}. This matrix is
written as the product of two matrices, $R$ and $T$, which account for radiative
and reconstruction effects, respectively. Also provided are 
variations on $T$ to enable the evaluation of systematic
uncertainties. The QED-Born level binning of 429 bins is finer than the
experimental binning. 

The $R$ matrices are $429 \times 429$ diagonal and correct
the Born-level cross sections for $O(\alpha)$ QED effects. 

A reconstruction-matrix element $T_{ij}$ gives the probability that
an event generated in $x$-$Q^2$ bin~$j$ is reconstructed in 
experimental bin~$i$ so that $T$ has a dimension of~$153 \times 429$.

The uncertainties that need to be accounted for in the ZEUS data are the following.
\begin{enumerate}
    \item The uncertainty in the luminosity of each of the
    \eplp\ and \emip\ data sets;
    \item The  uncertainties related to the imperfect understanding of detector
    effects such as acceptance, energy resolution, \xit{etc}.  These are completely correlated between the two data sets.
    Variational matrices $T'$ are provided for 8 sources
    of uncertainty in each of the \emip\ and \eplp\ data sets. 
\end{enumerate}

It turns out that the ZEUS luminosity 
uncertainty of 1.8\% is the most important source of uncertainty
in this analysis.  

From the matrices provided, we can construct for each source of systematic
error the deviation matrix
\[
  A' = R\,(T'-T)
\]
which corresponds to a one standard deviation uncertainty, assuming that
the systematic errors are Gaussian-distributed. Denoting the one standard deviation uncertainty on the luminosity by $\beta_0$, and including weight factors 
$\beta_0,\ldots,\beta_8$ in the set of model parameters $\theta$,
\Eq{eqn:nu_approx} can be written as, in matrix notation,
\begin{equation}
    \nu = \mathcal{L}\, 
    (1+ 0.018 \cdot \beta_0)
    \left( A + \sum_k \beta_k A'_k \right) \lambda, 
\end{equation}
where the sum runs over the 8 sources of systematic error and the $\beta$
parameters are taken to be normally distributed with unit width.

%------------------------------------------------------------------------------------------
\subsection{Simulated data}
To validate the analysis framework, we generated pseudo-data by computing
event predictions $\nu$ and event counts $n$ that are Poisson distributed with mean $\nu$.
With the input parameters given in \tabref{tab:inputpar}.
%-------
\begin{table}[htb]
\caption{Parameter values used in the data simulation.}
\begin{tabular*}{\linewidth}{@{\extracolsep{\fill}}ccccccccc}
\hline\hline
\multicolumn{9}{c}{$\ve{\Delta} \times 10^{3}$} \\ \hline
$u^V$ & $d^V$ & $g^V$ & $g^S$ & $2\bar{u}$ & $2\bar{d}$ & $2\bar{s}$ & $2\bar{c}$ & $2\bar{b}$ \\ \hline
228 & 104 & 249 & 249 & 104 & 52 & 10 & 5 & 0.5 \\
\hline
\end{tabular*}
\begin{tabular*}{\linewidth}{@{\extracolsep{\fill}}cccccccc}
$K_u$ & $K_d$ & \lamvg & \lamsg  & $K_g$ & \lamqb & \Kqb &\ve{\beta} \mbstrut{1.5em} \\
\hline
3.70  & 3.70  &  0.50  & $-$0.50 & 5.0   & $-$0.50 & 6.0 & 0 \\
\hline\hline
\end{tabular*}
\label{tab:inputpar}
\end{table}%
%------
The input parameters are chosen such that the simulated pseudo-data is similar to the actual ZEUS data. All $\beta$ parameters are set to zero so that there are no systematic biases in the simulated data. To study convergence, we have also generated pseudo-data sets with the ZEUS luminosity increased by a factor of~50 so that the likelihood tends towards a delta function. 
%
%--------------------------------------------------------------------------------------------
\subsection{Fits to the simulated data \label{se:modelfits}}

Unless otherwise stated, we use the priors
listed in \tabref{tab:stdpriors} in the fits\footnote{
   In a Bayesian analysis, parameters are not fitted in the usual sense. 
   In the context of this paper, a fitted (or free) parameter is one that has a prior
   \emph{distribution} assigned to it. Fixed parameters, on the other hand, have 
   single-valued priors, thereby reducing the dimension of the parameter space
   to be explored.
   } 
to the pseudo-data.
%----------
\begin{table}[hb]
\caption{Priors used in the analysis of the pseudo-data sets. There are 9
parameters in the vector \ve{\Delta} and 10 in \ve{\beta}. 
The normal distributions are truncated to the
range indicated, and their mean and standard deviation are given in brackets.}
\begin{ruledtabular}
\begin{tabular}{cll}
   & \textbf{Prior}  & \textbf{Range}\\
 \hline
$\ve{\Delta}$        & Dir(20,\;10,\;20,\;20,\;5,\;2.5,\;1.5,\;1.5,\;0.5) & $[0,1]$       \\
$K_u$                & Normal(3.5, 0.5)                             & $[1,6.5]$       \\
$K_d$                & Normal(3.5, 0.5)                             & $[1,6.5]$       \\
$\Sup{\lambda}{v}_g$ & Uniform                                      & $[0,1]$       \\
$\Sup{\lambda}{s}_g$ & Uniform                                      & $[-1,-0.1]$   \\
$K_g$                & Normal(4, 1.5)                               & $[1,8.5]$       \\
$\lambda_{\bar{q}}$  & Uniform                                      & $[-1,-0.1]$ \\
$K_{\bar{q}}$        & Normal(4, 1.5)                               & $[1,9.5]$     \\
$\ve{\beta}$        & Normal(0, 1)                                  & $[-5,5]$      \\
\end{tabular}
\end{ruledtabular}
\label{tab:stdpriors}
\end{table}%
The matrices for describing the transformation from QED
Born-level cross sections to the observed level are taken from the ZEUS
Collaboration~\cite{ZEUS:2020ddd} for all tests described in this paper.  
Tests were performed both with the systematic $\beta$ parameters fixed to zero,
or with them left free. Including the systematic parameters, we have a total of 
26 free parameters in our fit. The runtime, in this case, is $\sim$250k~samples/hour 
on a single core.

\begin{figure}[htb] 
    \centering
    \includegraphics[width=0.235\textwidth]{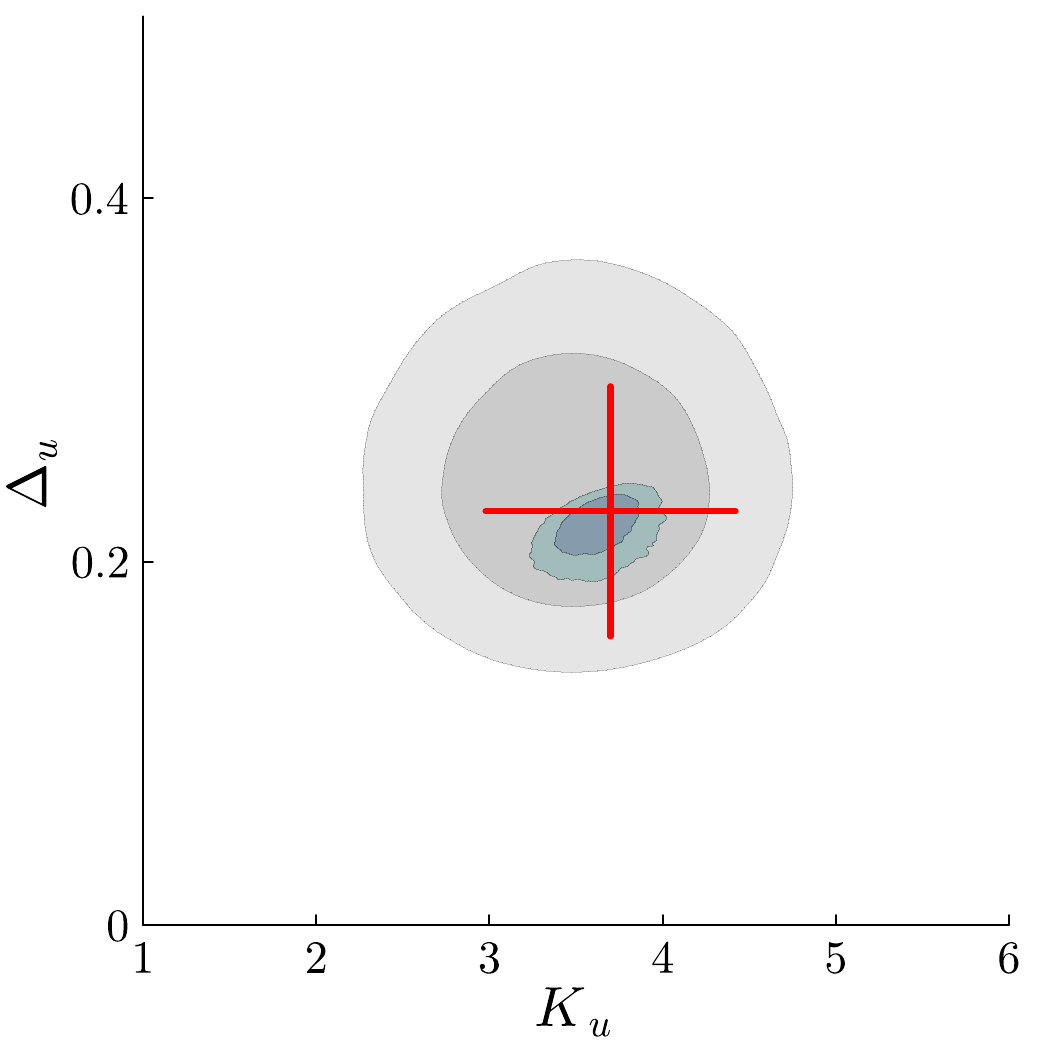}
    \includegraphics[width=0.235\textwidth]{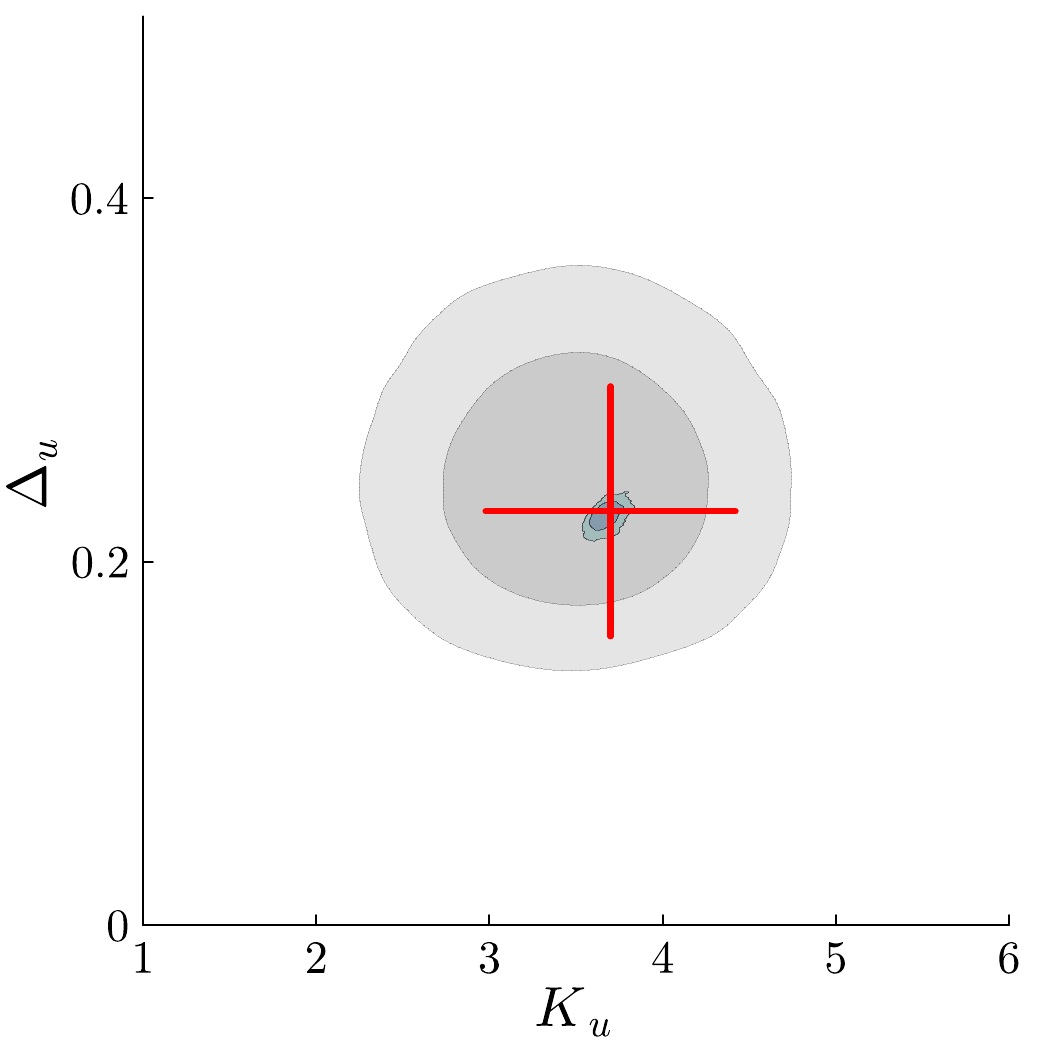}
    \llap{\raisebox{3.5cm}{%  move next graphics to top right corner
      \includegraphics[height=1cm]{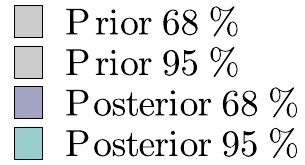}
    }}
    \includegraphics[width=0.235\textwidth]{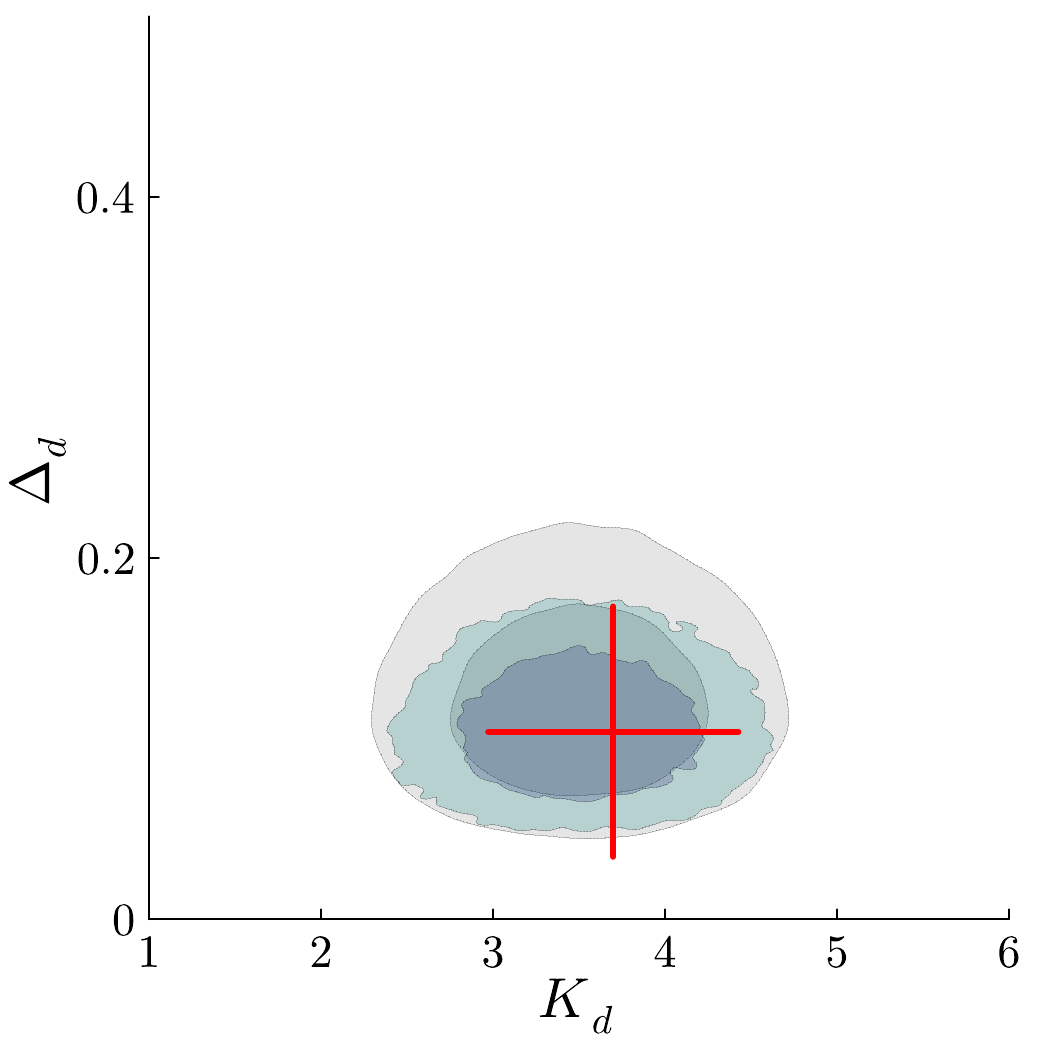}
    \includegraphics[width=0.235\textwidth]{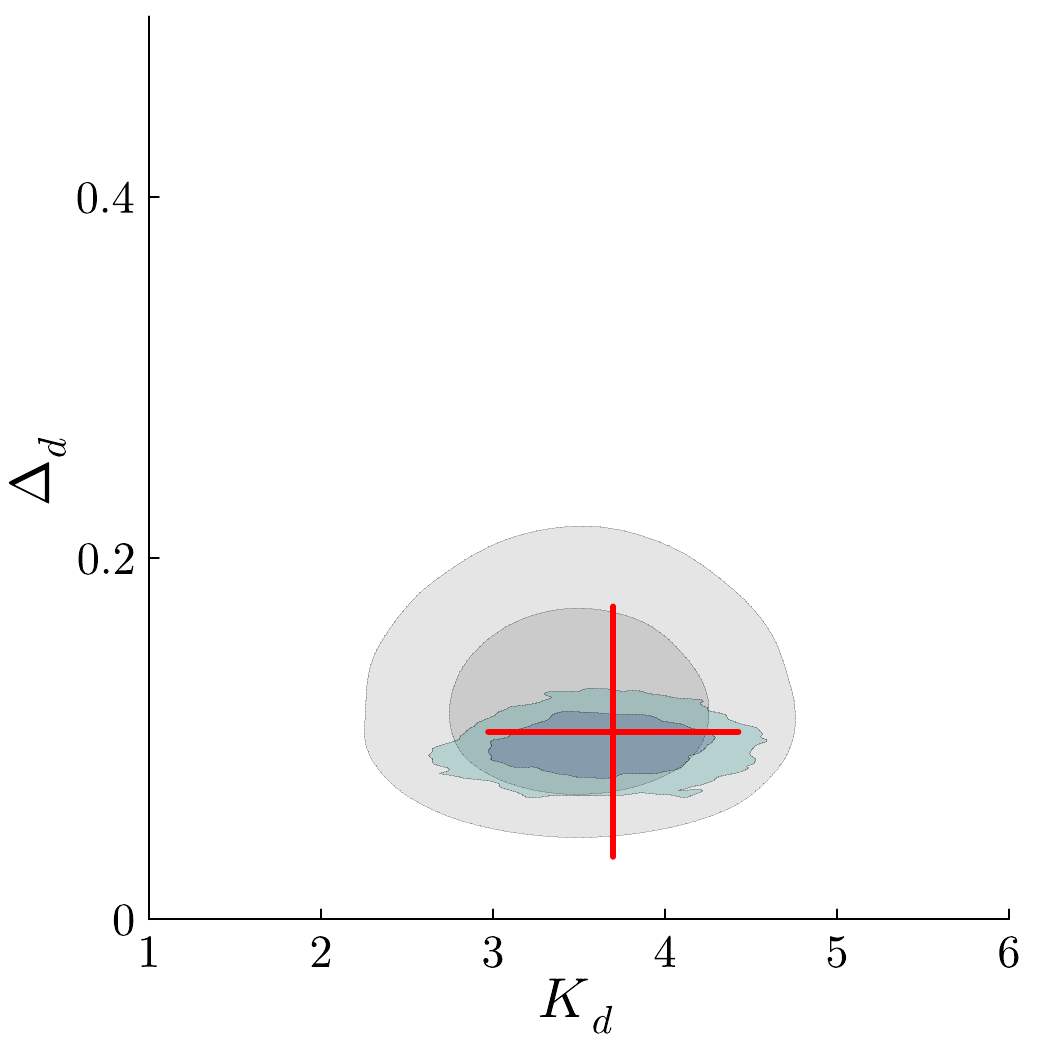}
    \caption{The prior and posterior probability distributions of $(\Delta_u,K_u)$ (upper)
             and $(\Delta_d,K_d)$ (lower) from the nominal (left) 
             and high-luminosity (right) pseudo-data sets. The red crosses indicate the known true 
             parameter values used in generating the pseudo-data.}
    \label{fig:KDcorrelation}
\end{figure}

\Fig{fig:KDcorrelation}
%---------------------
%----------------------
shows two-dimensional prior and posterior distributions for the parameters
$(\Delta_u, K_u)$ and $(\Delta_d, K_d)$ obtained from the nominal and high-luminosity
pseudo-data sets.
As we have already observed in~\cite{Aggarwal:2022cki}, a comparison of posterior
to prior shows a very significant knowledge update for the valence up-quark parameters, 
and less so for those of the valence down-quark.
The knowledge update on the parameters of the gluon and sea distributions (not shown)
is also rather limited, as is expected from an analysis of high-$x$ data.

Comparison of the left- and right-hand plots in \fig{fig:KDcorrelation} shows that the
posterior converges to the true values, indicating that there is no appreciable bias
introduced by our analysis procedure. We have verified that this is the case for all
fitted parameters, although convergence with increasing statistics is slow or absent
for those that are ill-constrained by the data.

\begin{figure}[htb] 
    \centering
    \includegraphics[width=0.23\textwidth]{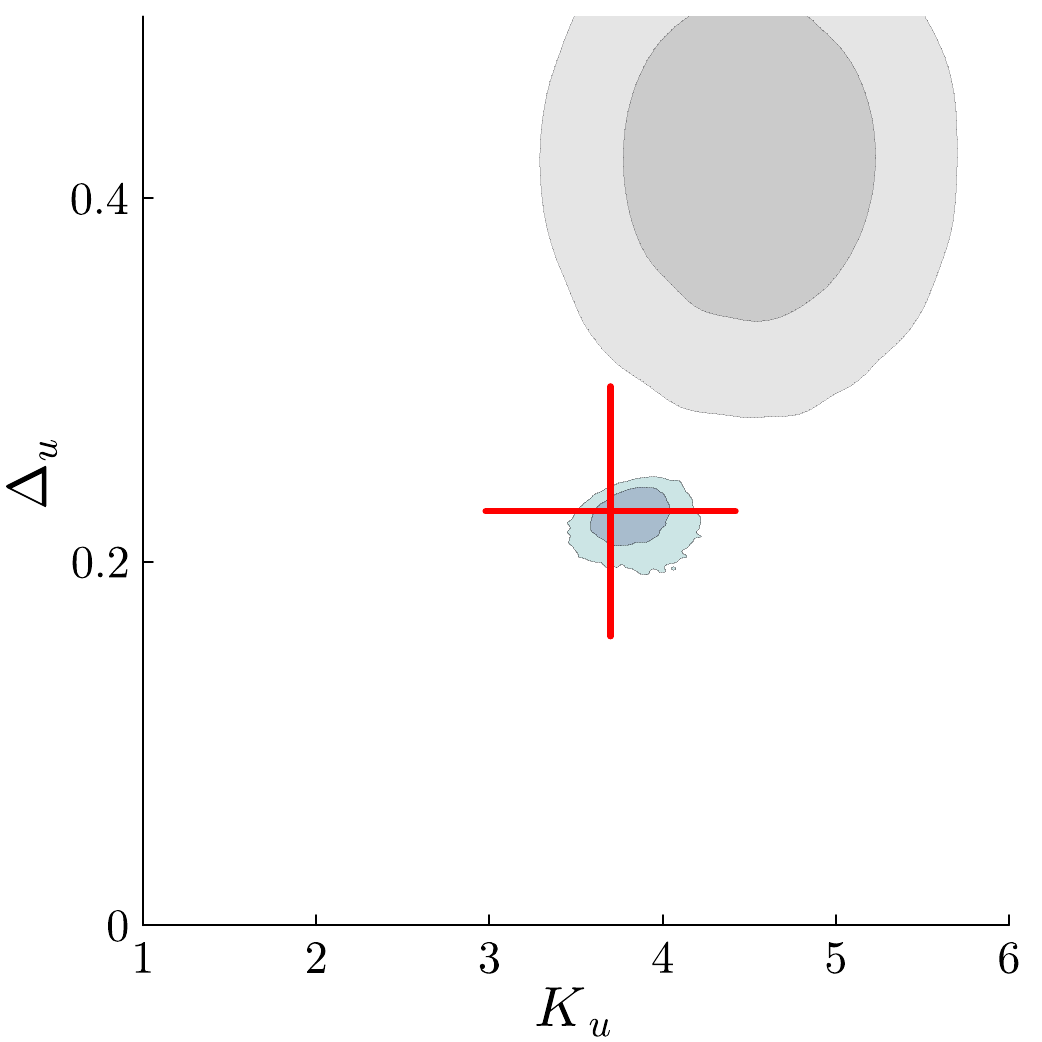}
    \includegraphics[width=0.23\textwidth]{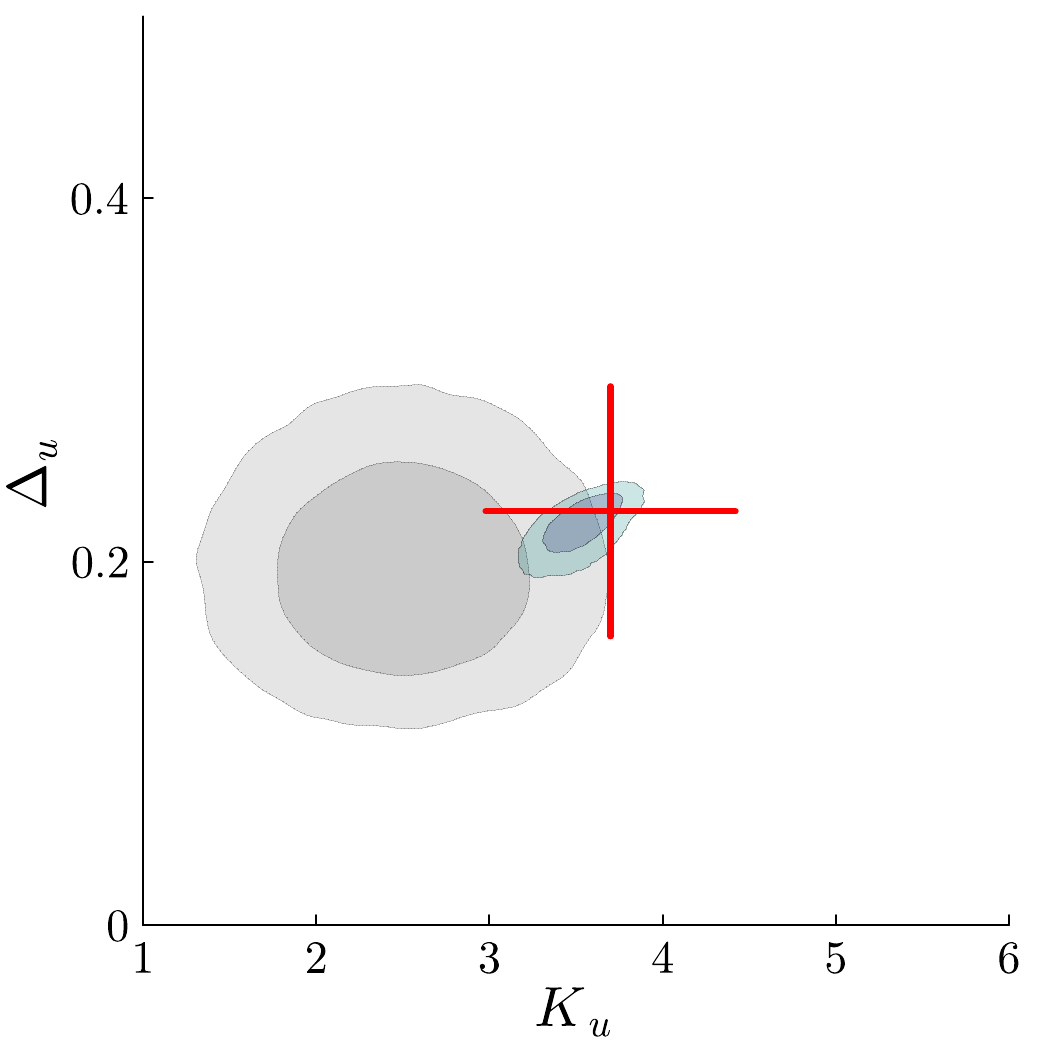}
    \caption{The posterior probability distributions of $(\Delta_u,K_u)$ 
             from two fits to the simulated pseudo-data with strongly biased priors (see text).
             True values are indicated
             by the red crosses. The legend is the same as that of \fig{fig:priorshift}.}
    \label{fig:priorshift}
\end{figure}

Parameters that are well-constrained by the data should be relatively insensitive to
the choice of prior. Many different data sets were analyzed using  shifted prior choices
to determine if these significantly affected the results. Two such tests for the
up-valence parameters are shown in~\fig{fig:priorshift}. 
%---------------------
%----------------------
In the left-hand plot, the momentum fraction $\Delta_u$ was strongly biased upward borrowing
momentum from the gluon while the prior for $K_u$ also had an upward bias.
In the right-hand plot, the gluon momentum was strongly biased upward, together with a
downward bias of $K_u$. In both cases, the posterior well reproduced the true values
of $\Delta_u$ and $K_u$.

\begin{figure}[htb] 
    \centering
    \includegraphics[width=0.475\textwidth]{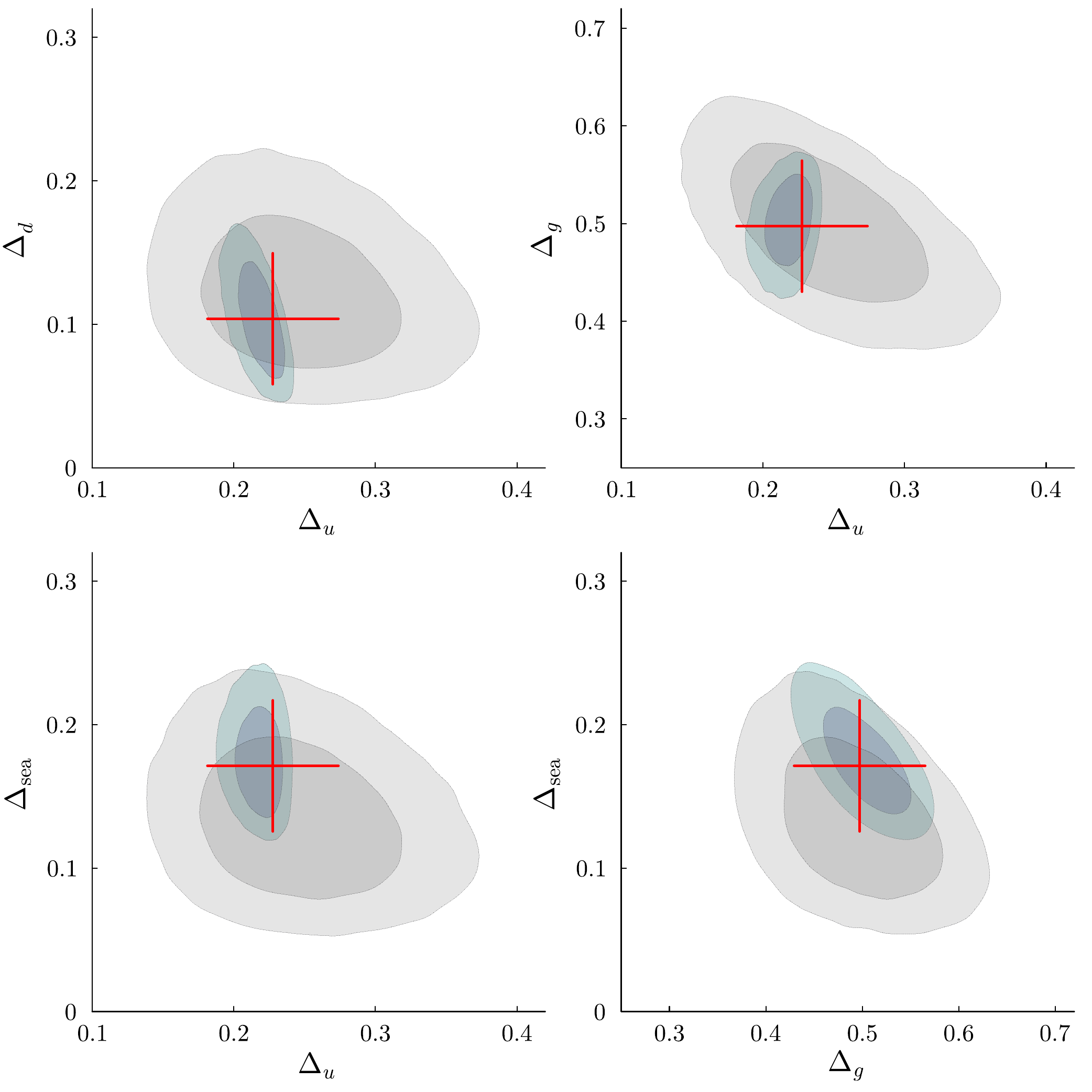}
    \caption{Correlations between parton momenta from the analysis  
             of the high-luminosity pseudo-data. True values are indicated by
             the red crosses. The legend is the same as that of \fig{fig:priorshift}.
             }     
    \label{fig:momenta}
\end{figure}

As a further example of the information made available by the full posterior
probability density we plot in~\fig{fig:momenta}
%---------------------
%----------------------
various correlations among the momentum components of the proton obtained from
the analysis of the high-luminosity data set. Here $\Delta_{\rm sea}$ is the fractional
momentum of the sea (anti-)quarks, summed over all quark flavors.
Again, it is seen that the true values are nicely reproduced and also that the
momentum carried by the valence up quark is very well constrained and
weakly sensitive to the other components.  The momenta carried by the sea quarks and the
gluon density are anti-correlated, as is to be expected since the sea is generated
from gluon splitting. For another correlation plot, we refer to \fig{fig:reportcorner}
in \secref{se:PDpackage}.

The results shown up to now were obtained from fits where the systematic parameters were
kept fixed to zero. Leaving these parameters free showed that the data hardly constrain
them and that they remain un-correlated, except for the \eplp\ and \emip\
luminosity parameters that show a strong correlation, as expected. 
However, a fit with free 
systematic parameters is always preferable
since that allows for error propagation by marginalizing the
posterior over them. However, with the pseudo-data sets studied here, this had a minor effect.

We mentioned in \secref{sec:priorconstraints} an alternative scheme where, instead of
fitting all parton momenta, those of the up and down valence quarks are fixed by fitting,
instead, their low-$x$ $\lambda$ shape parameters. In such a fit, we have removed
the priors on $\Delta_{u,d}$ and introduced those on $\lambda_{u,d}$ as normal
distributions of unit width centered at~$0.5$ and truncated to a range $[0.2,0.8]$.
It turned out that the results were very similar to those from the standard parameterization.
 
%---------------------
\begin{figure}[tbh]
    \centering
    \includegraphics[width=0.47\textwidth]{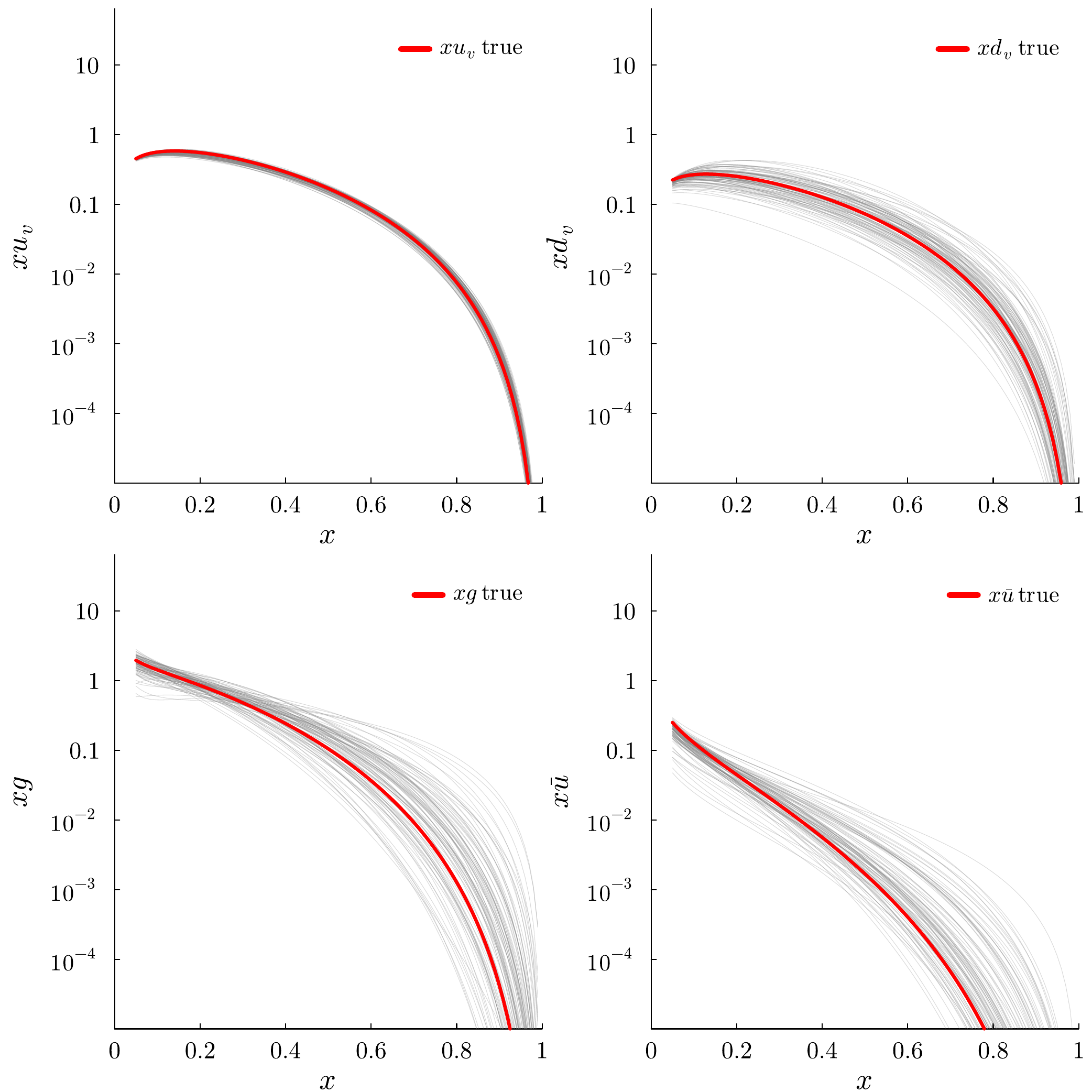}
    \caption{True parton momentum densities (red curves) compared to those
             computed from 100 samples of the posterior distribution (gray curves)
             from the simulated pseudo-data at nominal luminosity. The red lines represent the known true densities. 
             }     
    \label{fig:xfxplot}
\end{figure}
%----------------------
In \fig{fig:xfxplot} (note the vertical logarithmic scale) 
we show input parton distributions used in the generation of the
pseudo-data compared to those computed from 100 parameter sets randomly sampled from
the posterior distribution of the data set at nominal luminosity.  
Again, it is seen that the valence up-quark distribution is faithfully reproduced.
The valence down-quark distribution is poorly constrained and has a wide spread.
The general features of the gluon and up-antiquark distributions are correctly reproduced.
%
%-------------------------------------------------------------------------------------------
\section{Goodness-of-Fit Tests\label{se:GOFtests}}
A standard Bayesian analysis does not provide
goodness-of-fit tests that compare a single model to data. In fact, 
a single-model Bayesian analysis yields, as a result,
the posterior probability distribution of the model parameters, and no statement is made concerning the validity of the model itself.
To investigate model dependence, a choice is to be based on posterior probabilities calculated
for several alternative models~\cite{Beaujean:2011zza}. 
Such model selection is beyond the scope of this paper and will
be used in a future extension of the framework to investigate the sensitivity to different
parameterizations of the input \pdf[s].

Although not strictly Bayesian, it is possible, and maybe welcome, to provide a
goodness-of-fit test based on a single posterior probability distribution.
We consider two possibilities below.

\subsection{Posterior predictive check}
\label{sec:post_pred}
In a posterior predictive check, the actual data are compared to pseudo-data 
distributions generated from parameter
sets that are sampled from the posterior distribution.
The quality of the model can then be judged by observing good or bad overlap of the
pseudo- and observed data.

\begin{figure}[tbh]
    \centering
    \includegraphics[width=0.47\textwidth]{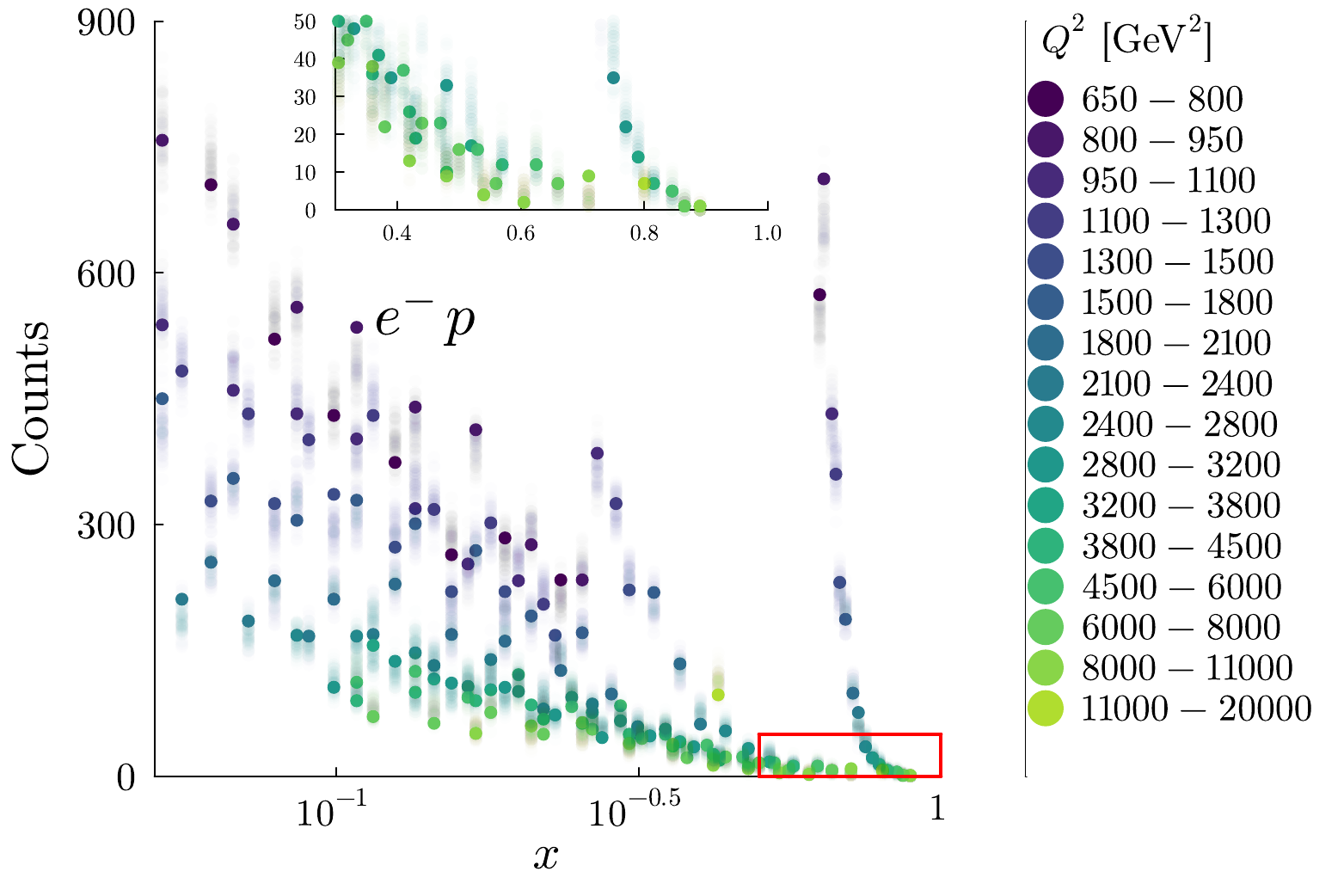}
    \caption{Event number distributions computed with parameters sampled
             from the posterior distribution (bands) compared
             to the simulated \emip\ scattering pseudo-data (dots).
             For clarity, the box at large $x$ is shown enlarged in the inset.
             }     
    \label{fig:dataspace}
\end{figure}

A graphical example of such a test is shown in \fig{fig:dataspace} 
%---------------------
%----------------------       
which shows good agreement. This is not surprising
since the data themselves were generated from a given parameter set, as is described in
\secref{se:validation}. Here the figure confirms that the input parameters are well-reproduced 
and that the analysis framework does not
introduce significant bias in the result.
%
%-------------------------------------------------------------------------------------------
%
\subsection{Posterior mode $\chi^2$ test}

\begin{figure}[tbh]
    \centering
    \includegraphics[width=0.4\textwidth]{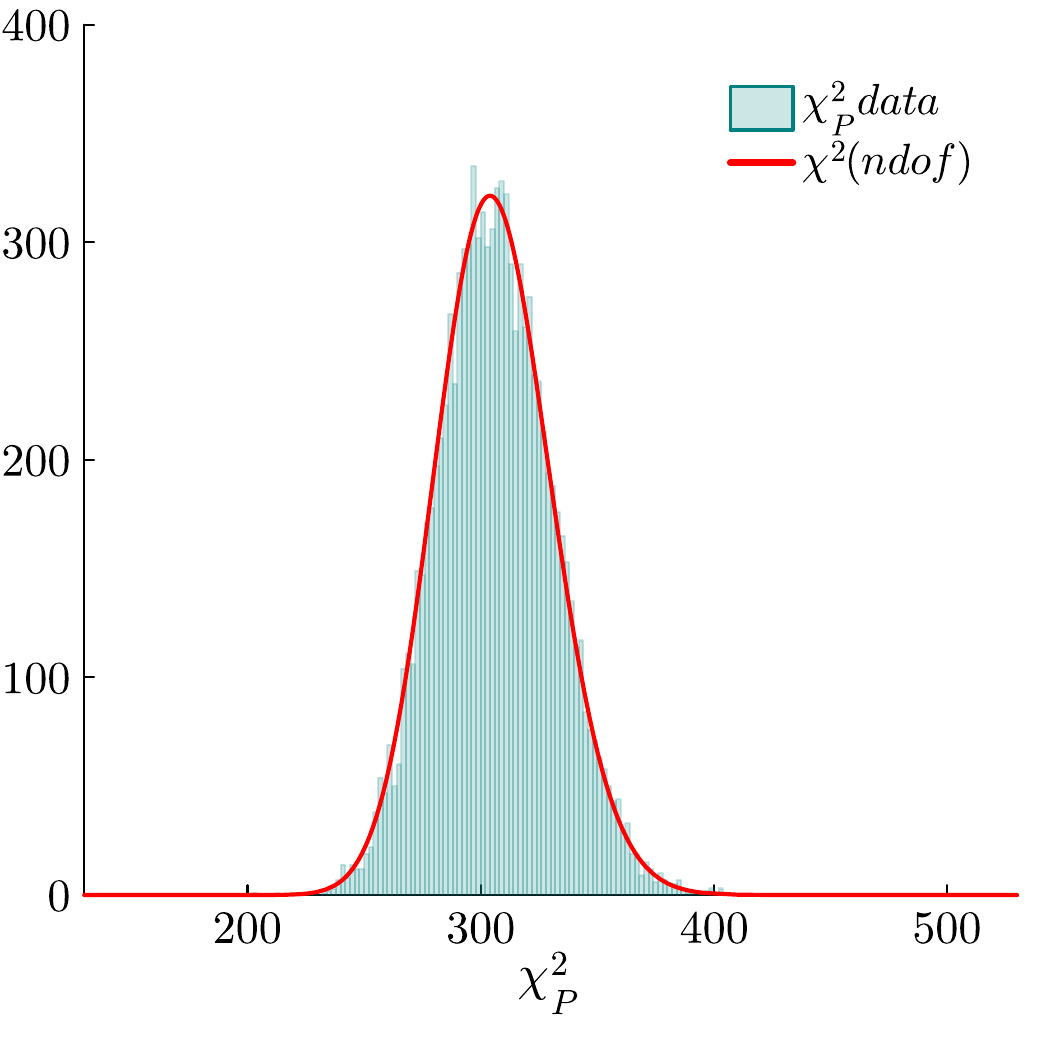}
    \caption{The $\chi^2$ distribution computed from the combined \eplp\ and \emip\ 
             analysis results. The curve shows the prediction for 306 degrees of freedom.}     
    \label{fig:chisq}
\end{figure}

A simple goodness-of-fit test is provided by computing the Pearson $\chi^2$, defined by
\[
 \chi^2_P = \sum_i \frac{(n_i - \nu_i)^2}{\nu_i},
\]
where the sum runs over all bins $i$ with observed event numbers $n_i$, 
and expected number of events $\nu_i$ as calculated from the 
posterior global mode parameter values.
In the presence of a sizable amount of sparsely populated or empty bins, $\chi^2_P$ does
not follow a standard $\chi^2$ distribution. In this case, it is better to calculate $\chi^2_P$ from a large set of simulated pseudo-data. In \fig{fig:chisq}
%---------------------
%----------------------
we show a histogram of the $\chi^2_P$ distribution obtained from many
\epmp\ pseudo-data sets that have $\nu_i$ fixed and $n_i$ Poisson distributed around $\nu$. 
From such histograms, it is straightforward to compute $p$-values by normalizing the histogram
and summing the bin contents above the observed value of $\chi^2_P$.
In our tests, we have found a fairly flat distribution of $p$-values, 
as expected. 

We have seen in our studies that the maximum posterior probability results do not
coincide with a minimum value of $\chi^2_P$.
This is not surprising since a maximum posterior is not necessarily
a maximum likelihood.
Nevertheless, the distribution in \fig{fig:chisq}
closely resembles that of a standard $\chi^2$ and,
as explained in ~\cite{Beaujean:2011zza, caldwellfrequentism}, 
small $p$-values can be understood
from a Bayesian perspective as implying that the model is open to improvement.
\vspace{0.5cm}

%-----------------------------------------------------------------------------------------
%
\section{Summary and Outlook}
We have developed a novel parton density analysis code that allows for a full Bayesian
posterior probability determination and also supports a forward modeling approach.
The open-source code has been thoroughly tested and is now available for distribution.
In this paper, we report on its structure, the technical developments that have been made
in realizing the code, and a series of validation tests that have been performed.  
We believe that the code is reliable, well-documented, and easy to use.

To date, the code has been used exclusively for the analysis of high-$x$ and high-$Q^2$ $e^{\pm}p$  deep inelastic scattering data.  
We look forward to also extending the analysis to other data sets, including those
reported as differential cross sections at the QED Born level, although such an analysis
cannot benefit from the statistical rigor offered by the forward modeling approach. 
In fact, the information needed to enable such an approach is often not made available 
in proper form by the experiments.

An important step in achieving this is to make the analysis code run much faster.
Here there is ample room for improvement by parallelizing computations
through threading or forking,
by improving the MCMC sampling efficiency and by speeding up the QCD evolution
of the \pdf[s].

We also intend to extend the framework to investigate more flexible \pdf\ parameterizations
using Bayesian model selection techniques.

\section*{Acknowledgements}

The authors would like to thank Richard Hildebrandt for extensive investigations of alternative PDF parameterizations, which we intend to report in a future publication. R. Aggarwal acknowledges the support of Savitribai Phule Pune University. F. Capel was employed by the Excellence Cluster ORIGINS, which is funded by the Deutsche Forschungsgemeinschaft (DFG, German Research Foundation) under Germany's Excellence Strategy - EXC-2094-390783311, during much of the project duration. The authors would like to dedicate this paper to the memory of Dr. Michiel Botje, who has contributed greatly to this work and will be deeply missed.

\bibliography{literatur.bib}

\end{document}